%% file: main.tex
\documentclass[osajnl,twocolumn,showpacs,superscriptaddress,nofootinbib,9pt]{revtex4-1}

\def\thesection{\arabic{section}}

\usepackage[english]{babel}

\usepackage{amsmath}
\usepackage{mathtools}
\usepackage{graphicx}
\usepackage{subfigure}
\usepackage[colorinlistoftodos]{todonotes}

\usepackage{hyperref}
\hypersetup{colorlinks=true,allcolors=blue}

\usepackage{amsmath}
\usepackage{amssymb}
\usepackage{centernot}
\usepackage{cancel}

\usepackage[normalem]{ulem}

\usepackage{enumitem}

\newcommand{\dd}{\mathrm{d}}
\newcommand{\LL}{\mathrm{L}}

\newcommand{\CC}{\mathrm{C}}

\newcommand{\pix}{\mathrm{pix}}

\begin{document}

\title{Photon noise correlations in millimeter-wave telescopes}

\author{Charles A. Hill}
\affiliation{Department of Physics, University of California, Berkeley, CA 94720, USA}
\affiliation{Physics Division, Lawrence Berkeley National Laboratory, Berkeley, CA 94720, USA}

\author{Akito Kusaka}
\email[]{akusaka@phys.s.u-tokyo.ac.jp}
\affiliation{Physics Division, Lawrence Berkeley National Laboratory, Berkeley, CA 94720, USA}
\affiliation{Department of Physics, University of Tokyo, Bunkyo-ku, Tokyo 113-0033, Japan}
\affiliation{Kavli IPMU, University of Tokyo, Kashiwa, Chiba 2778583, Japan}
\affiliation{Research Center for the Early Universe, University of Tokyo, Tokyo, 113-0033, Japan}


\begin{abstract}
Many modern millimeter and submillimeter (``mm-wave'') telescopes for astronomy are deploying more detectors by increasing detector pixel density, and with the rise of lithographed detector architectures and high-throughput readout techniques, it is becoming increasingly practical to overfill the focal plane. However, when the pixel pitch $p_{\rm pix}$ is small compared to the product of the wavelength $\lambda$ and the focal ratio $F$, or $p_{\mathrm{pix}} \lesssim 1.2 F \lambda$, the Bose term of the photon noise correlates between neighboring detector pixels due to the Hanbury Brown \& Twiss (HBT) effect.  When this HBT effect is non-negligible, the array-averaged sensitivity scales with detector count $N_{\mathrm{det}}$ less favorably than the uncorrelated limit of $N_{\mathrm{det}}^{-1/2}$. In this paper, we present a general prescription to calculate this HBT correlation based on a quantum optics formalism and extend it to polarization-sensitive detectors.  We then estimate the impact of HBT correlations on the sensitivity of a model mm-wave telescope and discuss the implications for focal-plane design.
\end{abstract}

\maketitle

\tableofcontents


\section{Introduction}
\label{sec:intro}
Modern millimeter and submillimeter (``mm-wave'') telescopes for astronomy are often limited by fluctuations in the background radiation. This is especially true for ground-based experiments where emission from the atmosphere and telescope are substantial. At high frequencies (e.g., optical wavelengths), the mode's mean occupation number $\bar{n} \ll 1$, and photon fluctuations are dominated by uncorrelated shot noise such that $\Delta \bar{n} \approx \sqrt{\bar{n}}$. At low frequencies (e.g., radio wavelengths), $\bar{n} \gg 1$ and photon fluctuations are dominated by the Bose term of the photon noise (``wave noise'') which correlates such that $\Delta \bar{n} \approx \bar{n}$. Millimeter wavelengths lie in a cross-over regime where $\bar{n} \sim 1$, making the calculation of array-averaged sensitivity in general nontrivial. 

In addition, many modern mm-wave telescopes, particularly those equipped with cryogenic bolometric detector arrays, are field-of-view-limited and therefore aim to increase detector count by increasing pixel density, which is typically cheaper than building more telescopes. In this high-pixel-density paradigm, it is possible to overfill the focal plane such that neighboring detectors sample the same spatial mode. As we will show, this mode sharing introduces photon-noise correlations when the pixel spacing $p_{\mathrm{pix}} < 1.2 F \lambda$, where $F$ is the effective focal ratio at the focal plane and $\lambda$ is the operational wavelength of interest. In this oversampled regime, photon noise correlations can have substantial impacts on the array-averaged sensitivity.

The theory of intensity correlations from incoherent sources has been studied extensively~\cite{fano_quantum_1961,glauber_coherent_1963,glauber_photon_1963,mandel_photon_1963,carter_coherence_1975,carter_coherence_1977}, and the phenomenon was experimentally demonstrated by Hanbury Brown and Twiss (HBT) via measurements of the angular diameter of distant astronomical sources \cite{brown_apparent_1952,brown_lxxiv_1954,brown_correlation_1956}.  The impact of HBT correlations on mm-wave telescopes is discussed by Padin~\cite{padin_mapping_2010}, where an empirical factor is introduced in an attempt to account for the corresponding sensitivity degradation.

In this paper, we present a prescription to estimate HBT correlations among detectors in millimeter- or submillimeter-wave telescopes based on a quantum optics formalism adopted from a circuit-based formalism for thermal photon correlations in quantum detectors developed by Zmuidzinas~\cite{zmuidzinas_thermal_2003}. We then extend this formalism to polarization-sensitive detectors and use it to calculate the impact of HBT correlations on the sensitivity of a model mm-wave telescope.

This paper is organized as follows.  In Sec.~{\ref{sec:theory}}, we review the theoretical foundations of our formalism based on Ref.~\cite{zmuidzinas_thermal_2003} and show how they relate to the HBT effect~\cite{brown_apparent_1952,brown_lxxiv_1954,brown_correlation_1956} and the van Cittert-Zernike theorem (VCZT)~\cite{vanCittert1934DieEbene,Zernike1938TheProblems}.  We also show the formalism's relation to the standard single-detector sensitivity calculations for a bolometer (e.g., see Ref.~\cite{hill_bolocalc_2018} and references therein).   Section~\ref{sec:optical_model} defines a model optical system for estimating the array-averaged sensitivity impact of HBT correlations.
In Sec.~\ref{sec:correlations}, we derive an expression for the intensity correlation using this model optical system.  Section~\ref{sec:sensitivity} discusses the impact of HBT correlations on the sensitivity of a telescope system with close-packed detectors on the focal plane. In Sec.~\ref{sec:implications_for_experiment_design}, we discuss the implications of the presented sensitivity optimization for the design of mm-wave detector arrays.  Finally, Sec.~\ref{sec:conclusion} presents our conclusions.


\section{Theoretical foundations}
\label{sec:theory}

In this section, we review the theory of photon-count statistics and reformulate them to the context of astronomical telescope systems. We first adopt the treatment of thermal photon correlations derived by Zmuidzinas~\cite{zmuidzinas_thermal_2003}, which uses the machinery of transmission lines and scattering matrices to calculate the propagation of quantum modes in a linear optical system.  We then apply this treatment to optical systems, where the optical equivalence theorem~\cite{sudarshan_equivalence_1963} allows us to equate the scattering matrix for quantum modes with the mode-mode coupling of classical waves (e.g., those obtained via physical optics calculations). We then show a few simple examples that relate this formalism to the the HBT effect, VCZT, and photon-noise calculations.


\subsection{Photon correlations} 
\label{sec:zmu_circuit}

\begin{figure}[btp]
    \centering
    \includegraphics[width=0.45\textwidth, trim=8cm 10cm 8cm 9cm, clip]{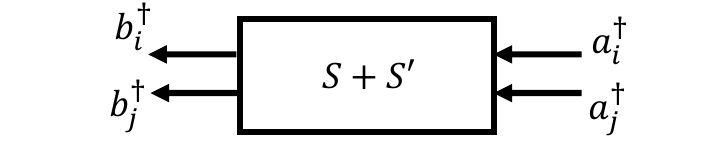}
    \caption{A schematic of the scattering matrix quantum circuit formalism. The creation operator for incoming modes is $a^{\dagger}$, while that of the outgoing modes is $b^{\dagger}$. The scattering matrix $S$ maps the input modes onto the output modes, while the noise matrix $S'$ calculates noise and loss within the system. $S$ and $S'$ are assumed to be orthogonal.}
    \label{fig:circuit_schematic}
\end{figure}

We first consider a linear, lossy network of $k=1, 2, \cdots$ input ports detected at an output port $i$.\footnote{There is no fundamental distinction between the inputs and outputs, and every port has both incoming and outgoing photons,
even though we will relate the input ports to optical input and the output ports to detectors.} Input modes enter the network along semi-infinite transmission lines via the photon creation operator $a_{k}^{\dagger}(\nu)$ and are mapped onto the outputs via the scattering matrix $S_{ik}$. Loss in the system is modeled by an orthogonal scattering matrix $S'_{ik}$, which governs the noise added between input mode $k$ and output mode $i$. Given this structure (Fig.~\ref{fig:circuit_schematic}), the creation operator $b_{i}^{\dagger}(\nu)$ at output $i$ and mode frequency $\nu$ is
\begin{equation}
    b_{i}^{\dagger}(\nu) = \sum_{k} S_{ik}(\nu) a_{k}^{\dagger}(\nu) + \sum_{k^{\prime}} S'_{ik^{\prime}}(\nu) a_{k^{\prime}}^{\dagger}(\nu) \, .
    \label{eq:scattering_creation_operators}
\end{equation}
As demonstrated in Eq.~(\ref{eq:scattering_creation_operators}), there is no fundamental distinction between the input/output ports and the lossy ports. Therefore, for simplicity, we hereafter absorb the scattering matrix for the lossy ports $S'_{ik}$ into $S_{ik}$ and treat both mechanisms via a single unified scattering matrix.

The two-photon expectation value at detector outputs $i$ and $j$ is given by
\begin{equation}
 \left< b_{i}^{\dagger} (\nu) b_{j} (\nu') \right>
 = \sum_{k} \sum_{m} S^{*}_{ik} (\nu) S_{jm} (\nu') \left< a_{k}^{\dagger} (\nu) a_{m} (\nu') \right> .
 \label{eq:two_photon_expectation_value}
\end{equation}
Here, the expectation values $\langle\cdots\rangle$ are taken over quantum-statistical mixed states, governed by the density matrix, and represent the quantum coherence of the photon modes at frequencies $\nu$ and $\nu'$. See Appendix~\ref{app:thermal_density_matrix} for further discussion regarding the thermal photon density matrix.

When the mixed states are in thermal equilibrium, which is a good approximation for the photon sources in the calculations that follow~\cite{zmuidzinas_thermal_2003}\footnote{The Kronecker delta $\delta_{km}$
in Eq.~(\ref{equ:ak_am_correlation}) indicates complete incoherence between the input source elements $k$ and $m$.
This is a good approximation for the applications discussed in this paper.
Further discussion on partial coherence of sources can be found in Appendix~\ref{app:partial_coherence}.},
\begin{equation}
\langle a_{k}^{\dagger} (\nu) a_{m} (\nu') \rangle = n(T_k, \nu) \, \delta_{km} \, \delta(\nu - \nu') \:,
\label{equ:ak_am_correlation}
\end{equation}
where $T_k$ is the temperature of port $k$ and
\begin{equation}
 n(T, \nu) \equiv \frac{1}{e^{h \nu / k_{\mathrm{B}} T} - 1}
 \label{eq:bose_einstein}
\end{equation}
is the mean occupation number at frequency $\nu$ of a blackbody at temperature $T$. We can write the two-photon output expectation value as
\begin{equation}
    \left< b_{i}^{\dagger} (\nu) b_{j} (\nu') \right> \equiv B_{ij}(\nu) \, \delta(\nu - \nu') \, ,
    \label{eq:Bij_definition}
\end{equation}
where $B_{ij}(\nu)$ is the quantum mutual intensity and satisfies
\begin{equation}
  \label{equ:mutual_intensity_vs_n}
    B_{ij}(\nu) = \sum_{k} S^{*}_{ik} (\nu) \, S_{jk} (\nu) \, n(T_k,\nu) \:.
\end{equation}
When calculated for a single detector $i$,
$B_{ii}(\nu)$ represents the mean occupation number
of the incoming photons at that detector.

Thermal detectors, which are commonly used in mm-wave applications, integrate photon power over time $\tau$ and sense mean intensity
\begin{equation}
    \langle  d_{i} \rangle = \frac{1}{\tau} \int_{0}^{\tau} \dd t \: \langle b_{i}^{\dagger}(t) b_{i}(t) \rangle \simeq \int_{\nu_{1}}^{\nu_{2}} \dd \nu \: h \nu B_{ii}(\nu) \, ,
    \label{eq:quant_bolo_detect_intensity}
\end{equation}
where we define the time-dependent operators as
\begin{equation}
\begin{split}
  b_{i}(t) & \equiv \int_{\nu_{1}}^{\nu_{2}} \dd \nu \: \exp \left[ 2\pi i \nu t \right] \: b_{i}(\nu) \sqrt{h \nu} \, , \\
  b_{i}^\dagger(t) & \equiv \int_{\nu_{1}}^{\nu_{2}} \dd \nu \: \exp \left[ - 2\pi i \nu t \right] \: b_{i}^\dagger(\nu) \sqrt{h \nu} \, .
  \label{eq:time_dependent_operators}
\end{split}
\end{equation}
Here, the integration limits are set by the detection bandwidth $\Delta \nu = \nu_{2} - \nu_{1}$, and the factors of $\sqrt{h\nu}$ arise due to power detection as opposed to photon counting.
In the context of free-space propagating modes, the operators $b_{i}(t)$ and $b_{i}^\dagger(t)$, with the factor $\sqrt{h\nu}$ inserted, can be regarded as electric field operators.
In typical detector readout configurations, the integration time $\tau$ can
be regarded as the inverse of the detector sampling rate.
The second equality in Eq.~(\ref{eq:quant_bolo_detect_intensity})
is a good approximation when $\tau \gg 1 / \Delta \nu$, which is often true in mm-wave experiments where $\tau \sim \mathcal{O}(10^{-2} $--$ 10^{-3} \; \mathrm{s})$ and $1 / \Delta \nu \sim \mathcal{O}(10^{-10} \; \mathrm{s})$.
Since the operators $b_{i}(t)$ and $b_{i}^\dagger(t)$ represent electric fields, a generalized form of Eq.~(\ref{eq:quant_bolo_detect_intensity}) corresponds to an expression of first-order coherence:
\begin{equation}
    \Gamma^{(1)}_{ij}
    =
    \frac{1}{\tau} \int_{0}^{\tau} \dd t \: 
    \langle b_{i}^{\dagger}(t) b_{j}(t) \rangle \simeq \int_{\nu_{1}}^{\nu_{2}} \dd \nu \: h \nu B_{ij}(\nu) \, .
    \label{eq:first_order_coherence}
\end{equation}
The normalized amplitude coherence $\gamma_{ij}$ can be written as
\begin{equation}
    \label{equ:gamma1_in_circuit_model}
    \gamma_{ij}
    \equiv \frac{\Gamma^{(1)}_{ij}}{\sqrt{\Gamma^{(1)}_{ii} \Gamma^{(1)}_{jj}}}
    \simeq \frac{B_{ij}(\bar{\nu})}{\sqrt{B_{ii}(\bar{\nu}) B_{jj}(\bar{\nu})}} \equiv \gamma_{ij}(\bar{\nu}) \:,
\end{equation}
where $\bar{\nu} \equiv (\nu_1 + \nu_2)/2$ is the mean frequency and the second equality is a good approximation
when the variation of $B_{ij}(\nu)$
is small within the detection band of $\nu_1 < \nu < \nu_2$.

Finally, the covariance for quantum thermal detectors $\sigma_{ij}^2 \equiv \langle \Delta d_{i} \Delta d_{j} \rangle = \langle d_{i} d_{j} \rangle - \langle d_{i} \rangle \langle d_{j} \rangle$ can be written as
\begin{equation}
    \sigma_{ij}^2
     \simeq \frac{1}{\tau} \int_{\nu_{1}}^{\nu_{2}} \dd \nu \, (h \nu)^{2} \left[ B_{ij}(\nu) \delta_{ij} + | B_{ij}(\nu) |^{2} \right]
    \label{eq:photon_count_covariance}
\end{equation}
as shown in Ref.~\cite{zmuidzinas_thermal_2003}.
The first term in the integrand of Eq.~(\ref{eq:photon_count_covariance}) represents uncorrelated shot noise, while the second term represents wave noise, which can correlate between output ports.  This second $|B_{ij}(\nu)|^{2}$ term is often referred to as the ``bunching term,'' as it quantifies the degree to which photon arrival times are correlated.  For convenience, we define the shot-noise and wave-noise parts of the covariance as
\begin{equation}
\label{equ:circuit_shot_wave}
\begin{split}
    \sigma_{ij,\mathrm{shot}}^2 & \equiv
        \frac{1}{\tau} \int_{\nu_{1}}^{\nu_{2}} \dd \nu \, (h \nu)^{2} B_{ij}(\nu) \delta_{ij} \:,
    \\
    \sigma_{ij,\mathrm{wave}}^2 & \equiv
      \frac{1}{\tau} \int_{\nu_{1}}^{\nu_{2}} \dd \nu \, (h \nu)^{2} | B_{ij}(\nu) |^{2} \: .
\end{split}
\end{equation}

Given the thermal detector covariance $\sigma_{ij}^{2}$ in Eq.~(\ref{eq:photon_count_covariance}), the (quantum) second-order coherence $\Gamma^{(2)}_{ij}$ can be defined as
\begin{equation}
    \Gamma^{(2)}_{ij}
    \equiv
    \tau \left< \Delta d_{i} \Delta d_{j} \right> = \tau \sigma^2_{ij} \:,
\end{equation}
where the factor $\tau$ comes from the fact that
$\left< \Delta d_{i} \Delta d_{j} \right>$
depends on the integration time $\tau$ (or the detector sampling rate $1/\tau$).
Therefore, the second-order coherence $\Gamma^{(2)}_{ij}$
represents the system's intrinsic degree of intensity coherence in $\mathrm{W^2 \cdot s}$ and is independent of integration time $\tau$.
These sampling-rate-independent fluctuations of detected photon power $\sqrt{\tau} \sigma_{ii}$
are equivalent to the detector's photon noise noise-equivalent power (NEP), as in Ref.~\cite{hill_bolocalc_2018}.

The normalized intensity coherence can be defined as
\begin{equation}
    \label{equ:gamma2_in_circuit_model}
    \gamma_{ij}^{(2)} = \frac{\tau \left< \Delta d_{i} \Delta d_{j} \right> }{\langle  d_{i} \rangle \langle  d_{j} \rangle / \sqrt{\Delta \bar{\nu}_i \Delta \bar{\nu}_j } }
    \simeq \frac{| B_{ij}(\bar{\nu})|^2}{B_{ii}(\bar{\nu}) B_{jj}(\bar{\nu})}
        \equiv \gamma_{ij}^{(2)} (\bar{\nu}) \:,
\end{equation}
with the detector bandwidth defined as
\begin{equation}
    \label{equ:bandwidth_definition}
    \Delta \bar{\nu}_i \equiv \frac{\int_{\nu_{1}}^{\nu_{2}} \dd \nu \, (h \nu)^{2} B^2_{ii}(\nu)}
    {\left[ \int_{\nu_{1}}^{\nu_{2}} \dd \nu \: h \nu B_{ii}(\nu) \right]^2}
    \simeq
    \frac{ \tau \, \sigma_{ii,\mathrm{wave}}^2 }{ \langle d_i \rangle^2 }
    \: .
\end{equation}
The second equality in Eq.~(\ref{equ:gamma2_in_circuit_model}) is a good approximation
when the variation of the integrand $| B_{ij}(\bar{\nu})|$ is small within the detection band $\nu_1 < \nu < \nu_2$.
The normalized intensity coherence corresponds to the correlation coefficient of the wave-noise covariance
$\gamma_{ij}^{(2)} \simeq \sigma^2_{ij, \mathrm{wave}} / \sigma_{ii, \mathrm{wave}} \, \sigma_{jj, \mathrm{wave}}$.
In other words, in the limit of a large occupation number $B_{ii}(\nu) \gg 1$,
the mean intensity $\langle d_i \rangle$ and its variance are related via the radiometer equation:
\begin{equation}
  \sqrt{\left< \Delta d_{i}^2 \right>} 
  = \frac{\langle  d_{i} \rangle}{\sqrt{\tau \, \Delta \bar{\nu}_i}} \:.
\end{equation}

The normalized intensity and amplitude coherences can be related as
\begin{equation}
    \label{eq:correl_1_vs_2}
    \gamma_{ij}^{(2)} \simeq \gamma_{ij}^{(2)} (\bar{\nu})
    = \left|\gamma_{ij} (\bar{\nu})\right|^2
    \: .
\end{equation}
This can also be derived for generic classical fields with complex Gaussian-random fluctuations (see, e.g., Ref.~\cite{Wolf2007IntroductionLight} and references therein).
For reasons described in Sec.~\ref{sec:simple_example_wo_pol}, we hereafter call
$\gamma_{ij} (\nu)$ and $\gamma_{ij}^{(2)}(\nu)$ the VCZT and HBT coefficients, respectively.

The intensity coherence $\gamma_{ij}^{(2)}$ is not affected by decoherence. Decoherence is present, or the complex phase of $\gamma_{ij}(\nu)$ rotates over the detection band such that $|\gamma_{ij}| < |\gamma_{ij}(\bar{\nu})|$, when the path-length difference between the light source and detectors $(i,j)$ is larger than the inverse of the detection bandwidth $|R_i - R_j| \gtrsim c/\Delta \nu$. However, $\gamma_{ij}^{(2)}$ is a real-valued positive quantity, and thus such an effect is nonexistent.\footnote{In other words, while the approximation in Eq.~(\ref{equ:gamma1_in_circuit_model}) neglects decoherence, those in Eqs.~(\ref{equ:gamma2_in_circuit_model}) and (\ref{eq:correl_1_vs_2}) do not, and the second equality in Eq.~(\ref{eq:correl_1_vs_2}) is exact.}  The intensity signal would still de-correlate if $|R_i - R_j|$ were larger than $c \tau$, but
for typical mm-wave experiments,
$\tau \sim \mathcal{O}(10^{-2} $--$ 10^{-3} \; \mathrm{s})$ and detectors are arranged such that $|R_i - R_j| \lesssim 10^{-3} \, \mathrm{m}$, leading to $|R_i - R_j| \ll c \tau$.
Therefore, de-correlation can be safely ignored.


\subsection{Simple example without polarization}
\label{sec:simple_example_wo_pol}

We now apply the formalism in Sec.~\ref{sec:zmu_circuit} to mm-wave optical systems.
As noted in the Appendix~A of Ref.~\cite{zmuidzinas_thermal_2003}, the quantum circuit treatment is readily applicable to free-space propagating waves, as the optical equivalence theorem~\cite{sudarshan_equivalence_1963} allows us to equate the scattering matrix to mode-mode coupling coefficients of classical electromagnetic wave amplitudes.

\begin{figure}[btp]
    \centering
    \includegraphics[width=0.48\textwidth, trim=6cm 7cm 3cm 1.5cm, clip]{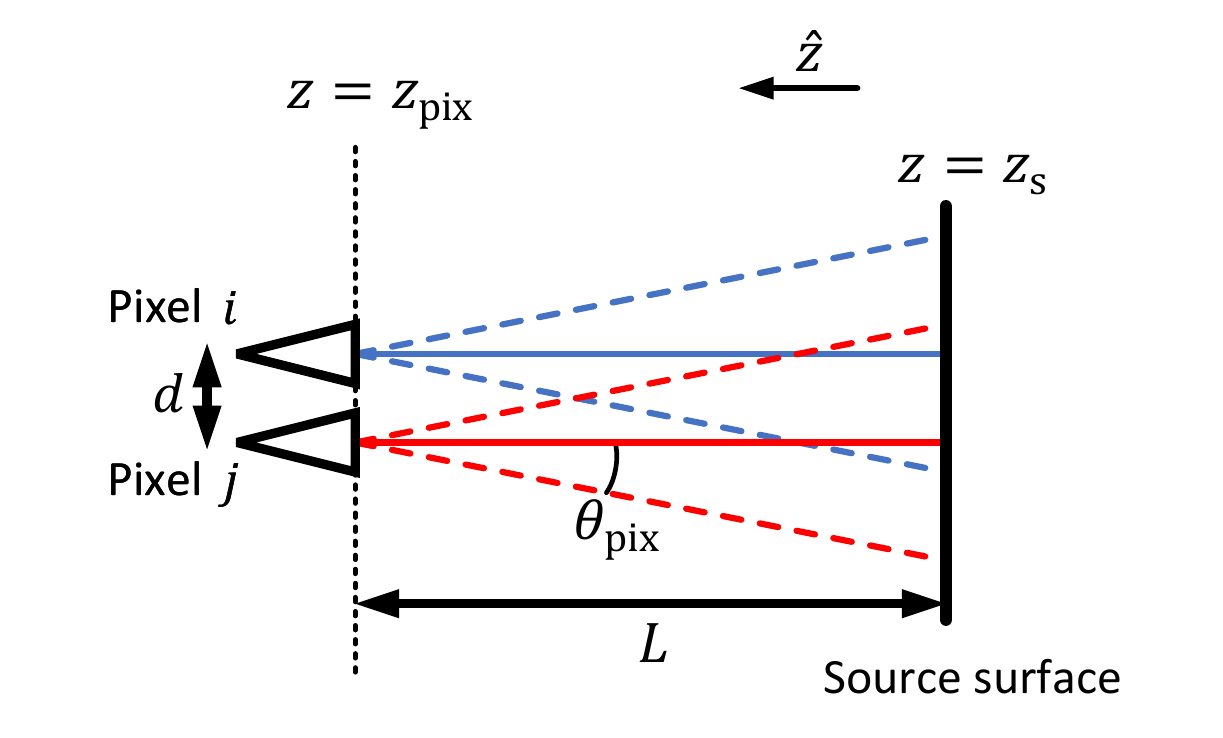}
    \caption{A simple example where two detectors ($i$ and $j$) on the $z=z_{\rm pix}$ plane observe light from a surface source on the $z = z_{\rm s}$ plane.  The direction of the rays are defined by spherical coordinate $(\theta_{\mathrm{pix}}, \phi_{\mathrm{pix}})$.}
    \label{fig:simplest_HBT}
\end{figure}

First, we consider the simplest case shown in Fig.~\ref{fig:simplest_HBT} with two identical planar detectors at $z = z_{\mathrm{pix}}$ and a far-field planar source at $z=z_s$. We assume that $|z_s-z_{\rm pix}| \equiv L \gg 2D_{\rm pix}^2 / \lambda$, where $D_{\rm pix}$ is each detector's aperture diameter and $\lambda \equiv c/\nu$ is the free-space electromagnetic wavelength, and we assume that the source is thermal with 100\% emissivity. Given the classical-wave amplitude of the electric field $E_i$ detected by detector $i$, the
partial field amplitude $\Delta E_i$ from an infinitesimal area of the planar source $\Delta s_k$ can be written as
\begin{equation}
    \Delta E_i = \mathcal{C} \, G(\theta_{i, \pix}, \phi_{i, \pix}) \,
    \sqrt{\cos \theta_{i, \pix}} \, \frac{e^{2\pi i \nu R_i / c}}{R_i} \, \Delta s_k
    \:,
    \label{equ:classical_coupling_in_simple_spherical}
\end{equation}
where $(\theta_{i, \pix}, \phi_{i, \pix})$ is the polar coordinate of the line between the detector and
the infinitesimal source area, $R_i$ is the distance between the detector and the infinitesimal area, $G(\theta_{i, \pix}, \phi_{i, \pix})$ is the detector's angular response function, $\mathcal{C}$ is a constant, and $\sqrt{\cos \theta_{i, \pix}}$ is a Lambertian factor.  While not explicitly written, $\mathcal{C}$ and $G(\cdots)$ may depend on frequency $\nu$.

As noted previously, the optical equivalence theorem allows us to relate the right-hand side of Eq.~(\ref{equ:classical_coupling_in_simple_spherical}) with the scattering matrix $S_{ik}(\nu)$. Thus, following Eq.~(\ref{equ:mutual_intensity_vs_n}), the mutual intensity can be calculated as
\begin{widetext}
\begin{equation}
    B^{\rm np}_{ij}(\nu) = |\mathcal{C}|^2 \iint_\sigma \!\! \dd^2 s_k \,
      G^*(\theta_{i, \pix}, \phi_{i, \pix}) G(\theta_{j, \pix}, \phi_{j, \pix})
      \sqrt{\cos \theta_{i, \pix}\cdot \cos \theta_{j, \pix}}
      \frac{e^{2\pi i \nu (R_j - R_i) / c}}{R_i R_j}
      n(T_k,\nu)
      \: ,
      \label{equ:mutual_intensity_nonpolarized}
\end{equation}
\end{widetext}
where $T_k$ is the temperature of the infinitesimal source $\Delta s_k$,
the superscript ``{\rm np}'' denotes ``no polarization,''
and the integral is over the source surface $\sigma$.
This form clarifies that Eqs.~(\ref{equ:mutual_intensity_vs_n}), 
(\ref{eq:first_order_coherence}), and (\ref{equ:gamma1_in_circuit_model}) can be regarded as a representation of the van Cittert-Zernicke theorem (VCZT)~\cite{vanCittert1934DieEbene,Zernike1938TheProblems} for generalized mode-coupling configurations specialized for thermal sources.

For a very distant circular source (such as a star) at $x=y=0$ and with temperature $T$, the mutual intensity can be simplified to
\begin{equation}
    B^{\rm np}_{ij}(\nu) \simeq \frac{|\mathcal{C} \, G(0,0)|^2}{L^2} n(T,\nu) \iint_\sigma \dd^2 s_k \,
    e^{2\pi i \nu (R_j - R_i) / c}
    \:.
    \label{equ:simple_mutual_intensity_vczt}
\end{equation}
In this case, the normalized spectral amplitude coherence $\gamma_{ij}(\nu)$ can be expressed as
\begin{equation}
    \gamma^{\rm np}_{ij}(\nu)
    =
    \frac{B^{\rm np}_{ij}(\nu)}{\sqrt{B^{\rm np}_{ii}(\nu) \, B^{\rm np}_{jj}(\nu)}}
    \simeq
    \frac{J_1 (2\pi \nu \, \theta_s \, p_{\rm pix} / c)}{\pi \nu \, \theta_s \, p_{\rm pix} / c} \: ,
    \label{equ:no_polarization_amplitude_coherence}
\end{equation}
where $J_1(z)$ is a Bessel function of the first kind of order one, $\theta_s$ is the angular radius of the source, $p_{\rm pix}$ is the distance between detectors $i$ and $j$, and
the normalized intensity coherence $\gamma^{(2)}_{ij} \simeq |\gamma^{\rm np}_{ij}(\bar{\nu})|^2$ reproduces the observation by Hanbury Brown and Twiss (HBT)~\cite{brown_correlation_1956}.

\subsection{Simple example with polarization}
\label{sec:simple_with_pol}

Building on Sec.~\ref{sec:simple_example_wo_pol}, we now introduce the polarization degree of freedom.
We adopt the same geometry as Fig.~\ref{fig:simplest_HBT} and
assume that each pixel is an ideal dual-polarization detector equipped with two orthogonal linear polarimeters.
When decomposing the propagating field into two polarization degrees of freedom, it is convenient to adopt the Ludwig-3 basis set~\cite{ludwig_definition_1973}
\begin{equation}
\begin{split}
    \hat{e}_1^{L3}(\theta, \phi) & \equiv \hat{\theta} \cos \phi  - \hat{\phi} \sin \phi  \: , \\
    \hat{e}_2^{L3}(\theta, \phi) & \equiv \hat{\theta} \sin \phi  + \hat{\phi} \cos \phi \: ,
\end{split}
    \label{eq:ludwig_polariztion}
\end{equation}
where $(\theta, \phi)$ defines the wave propagation direction in polar coordinates, and 
$\hat{\theta}$ and $\hat{\phi}$ are unit vectors in the direction of $(\theta, \phi)$.

Each detector pixel $i$ comprises two detectors $i_1$ and $i_2$ with polarization angles $\psi_i$ and $\psi_i + \pi/2$, respectively.
The angle $\psi_i$ is defined in the $x$-$y$ plane such that $\psi_i = 0$ when the polarization angle of detector $i_1$ is along the $x$-axis. Assuming ideal polarimetry,\footnote{Ideal polarimetry is often
characterized as having low cross polarization.
See, e.g., Refs.~\cite{kildal_2000,Stutzman_2000,zeng2010,Kusaka2014} for further discussion.
} detector $i_1$ only responds to the propagating electric field with polarization direction
\begin{equation}
    \hat{e}_{i1} = \hat{e}_1^{L3} \cos \psi_i  + \hat{e}_2^{L3} \sin \psi_i \: .
    \label{equ:ideal_detector}
\end{equation}
In other words, in the reverse-time sense,  the electric field emitted from detector $i_1$ has polarization direction $\hat{e}_{i1}$ as a function of $(\theta,\phi)$ in the far field.
Similarly, detector $i_2$ only responds to polarization direction
\begin{equation}
    \hat{e}_{i2} = - \hat{e}_1^{L3} \sin \psi_i  + \hat{e}_2^{L3} \cos \psi_i \: .
    \label{equ:ideal_detector2}
\end{equation}

The surface source can also be decomposed into two polarization degrees of freedom, denoted by $k_1$ and $k_2$.  This decomposition is arbitrary as long as $k_1$ and $k_2$ are independent and orthogonal Gaussian-random emitters, which is the case for unpolarized thermal sources. We therefore assume for simplicity that sources $k_1$ and $k_2$ emit with polarization  $\hat{e}_1^{L3}$ and $\hat{e}_2^{L3}$, respectively. 

The coupling between detector $i_1$ and source $k_1$ is similar to Eq.~(\ref{equ:classical_coupling_in_simple_spherical}), except for an additional polarization overlap factor $(\hat{e}_{i1} \cdot \hat{e}_1^{L3})$:
\begin{equation}
\begin{split}
    \Delta E_{i1}  =  (\hat{e}_{i1} \cdot \hat{e}_1^{L3}) \,
    \mathcal{C} & \, G(\theta_{i, \pix}, \phi_{i, \pix}) \, \\
    & \cdot \: \sqrt{\cos \theta_{i, \pix}} \, \frac{e^{2\pi i \nu R_i / c}}{R_i} \, \Delta s_{k1}
    \:.
\end{split}
    \label{equ:classical_coupling_in_simple_spherical_pp}
\end{equation}
Taking the coefficient on the right-hand side as scattering matrix element $S_{ik}(\nu)$ and following Eq.~(\ref{equ:mutual_intensity_vs_n}), 
the mutual intensity between detectors $i_1$ and $j_1$ can be expressed as
\begin{equation}
\begin{split}
    B_{i1, j1}(\nu) & = \sum_{p=1,2} (\hat{e}_{i1} \cdot \hat{e}_p^{L3})
    (\hat{e}_{j1} \cdot \hat{e}_p^{L3})
    B_{ij}^{\rm np}(\nu)
    \\
    & = \cos (\psi_i - \psi_j) B_{ij}^{\rm np}(\nu) \:,
\end{split}
    \label{equ:mutual_intensity_pol1}
\end{equation}
where $B_{ij}^{\rm np}(\nu)$ is the ``no polarization'' mutual intensity in Eq.~(\ref{equ:mutual_intensity_nonpolarized}).
The mutual intensity for other detector and polarization combinations can be calculated similarly, and the resulting amplitude coherences are
\begin{equation}
    \begin{split}
        \gamma_{i1, j1}(\nu) & = \gamma_{i2, j2}(\nu)
        = \cos (\psi_i - \psi_j) \, \gamma_{ij}^{\rm np}(\nu) \:,
        \\
        \gamma_{i1, j2}(\nu) & = - \gamma_{i2, j1}(\nu)
        = \sin (\psi_i - \psi_j) \, \gamma_{ij}^{\rm np}(\nu) \:,
        \\
        \gamma_{i1, i2}(\nu) & = \gamma_{j1, j2}(\nu) = 0 \: ,
    \end{split}
    \label{equ:vczt_with_pol}
\end{equation}
where $\gamma_{ij}^{\rm np}(\nu)$ is defined in Eq.~(\ref{equ:no_polarization_amplitude_coherence}).

It is convenient to calculate the covariance of the Stokes parameters, which are defined as the difference in power measured by two orthogonal polarimeters in a single detector pixel,  $Q_i \equiv (d_{i1} - d_{i2})/2$.  The Stokes-parameter covariance between two pixels can be calculated as
\begin{equation}
    \langle \Delta Q_i \Delta Q_j \rangle
    =
    \frac{1}{2} \cos [2(\psi_i - \psi_j)] \left( \sigma_{ij}^{\rm np}\right)^2
    \:,
    \label{equ:qq_covariance}
\end{equation}
where $\sigma_{ij}^{\rm np}$ is defined by substituting
$B_{ij}^{\rm np}(\nu)$ into Eq.~(\ref{eq:photon_count_covariance}).
We can then define the Stokes $Q$ normalized coherence
analogously to the intensity coherence in Eq.~(\ref{equ:gamma2_in_circuit_model})
\begin{equation}
    \gamma_{ij}^{Q, (2)}
    \equiv 
    \frac{\tau \left< \Delta Q_{i} \Delta Q_{j} \right> }{\langle  I_{i} \rangle \langle  I_{j} \rangle / \sqrt{\Delta \bar{\nu}_i \Delta \bar{\nu}_j } }
    = \cos [2(\psi_i - \psi_j)] \, \gamma_{ij}^{{\rm np}, (2)} \:,
    \label{equ:qq_correlation}
\end{equation}
where $I_i \equiv (d_{i1} + d_{i2})/2$ is the Stokes intensity,
and $\gamma_{ij}^{{\rm np}, (2)}$ is the ``no polarization'' intensity coherence defined via Eq.~(\ref{equ:gamma2_in_circuit_model}).  As expected, $\gamma_{ij}^{Q, (2)}$
corresponds to the correlation coefficient of the wave noise components of $Q_i$ and $Q_j$.

As previously mentioned and further discussed in Sec.~\ref{sec:optical_model}, the form in Eq.~(\ref{equ:qq_correlation}) that only introduces a $\cos [2(\psi_i - \psi_j)]$ factor to the ``no polarization'' case is general as long as the telescope and detector conform to the assumption of ideal polarimetry.  It is worth noting that, according to Eq.~(\ref{equ:qq_correlation}), the focal plane can be designed to minimize Stokes $Q$ correlation by assigning $\pm 45^\circ$ polarization angles to neighboring pixels.


\subsection{Detector photon noise}
\label{sec:photon_noise_correlations}

We now relate the formalism in Sec.~\ref{sec:zmu_circuit} to the forms of photon noise for bolometric detectors often seen in the literature.

\begin{figure}[tbp]
    \centering
    \includegraphics[width=0.45\textwidth, trim=3.1cm 5cm 0.5cm 3cm, clip]{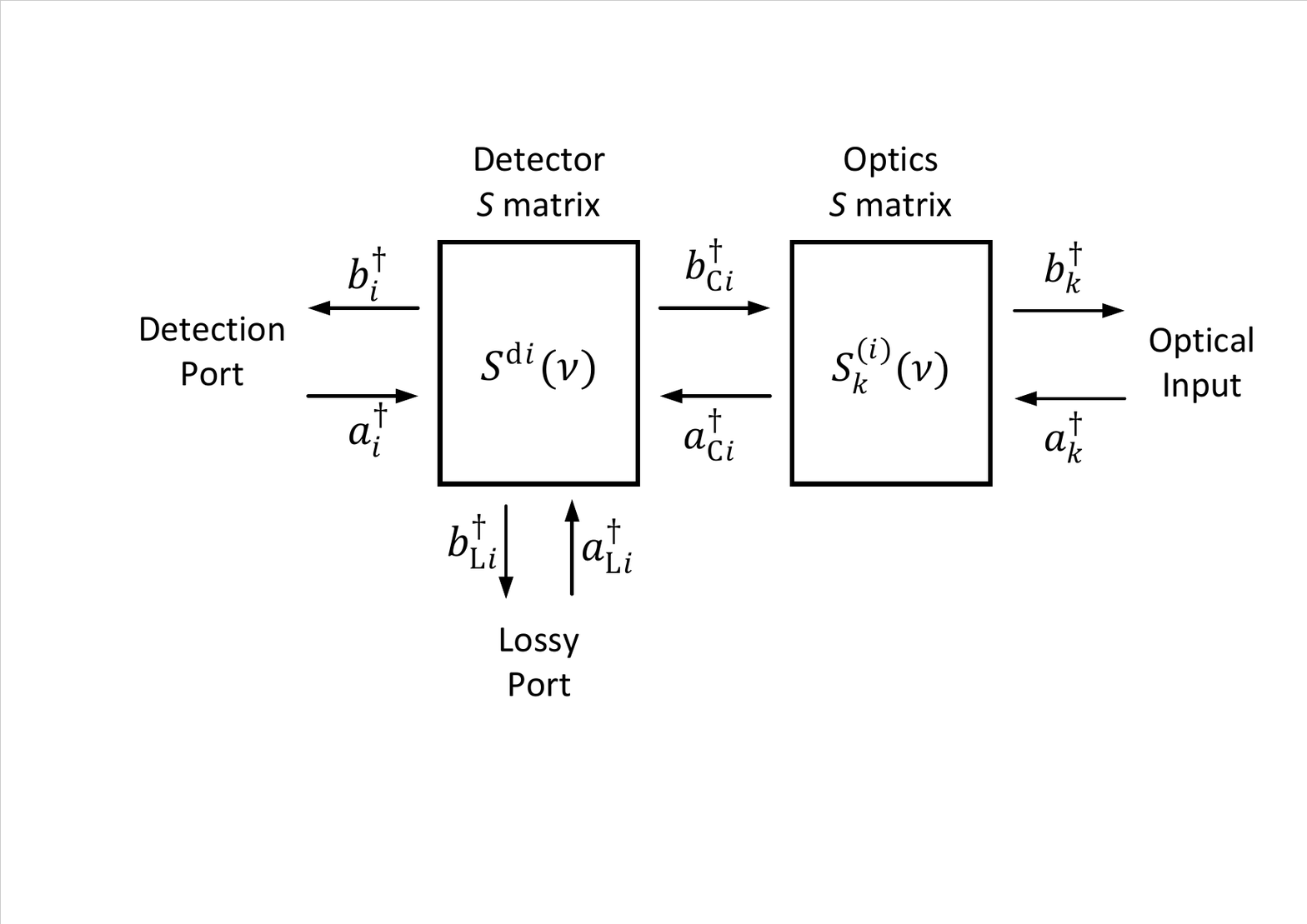}
    \caption{A schematic representing the detector scattering matrix $S^{\dd i}(\nu)$ and optics scattering matrix $S^{(i)}_k(\nu)$ for detector $i$.  The detector scattering matrix consists of three ports: the detection port $i$, the lossy port $\LL i$, and the optical coupling port $\CC i$.}
    \label{fig:detector_s_matrix}
\end{figure}

It is convenient to decompose the scattering matrix $S_{ik}$ between photon source $k$ and detector $i$ into that of the detector optics $S^{\dd i}_{ik}$ and that of the telescope optics that couple to detector $i$, $S^{(i)}_{k}$.  The former includes the detector's quantum efficiency (see below), and the latter includes coupling to the atmosphere, lossy optical elements, and any other detectable photon sources. The detector-only scattering matrix can be written as a simple three-port system as shown in Fig.~\ref{fig:detector_s_matrix}, where $i$, $\LL i$, and $\CC i$ represent the detection, loss, and optical-coupling ports, respectively.  The three corresponding scattering matrix elements, $S^{\dd i}_{i,i}$, $S^{\dd i}_{i,\LL i}$, and $S^{\dd i}_{i,\CC i}$, can be related to the reflection, loss, and transmission while satisfying the normalization
\begin{equation}
    |S^{\dd i}_{i,i}(\nu)|^2 + |S^{\dd i}_{i,\LL i}(\nu)|^2 + |S^{\dd i}_{i,\CC i}(\nu)|^2 = 1 \: .
\end{equation}
We define the quantum efficiency of detector $i$ as
\begin{equation}
    \eta_i(\nu) \equiv |S^{\dd i}_{i,\CC i}(\nu)|^2 = 1 - |S^{\dd i}_{i,i}(\nu)|^2 - |S^{\dd i}_{i,\LL i}(\nu)|^2 \: ,
    \label{equ:det_quantum_efficiency}
\end{equation}
which includes both the detector's efficiency and its spectral response.

For bolometric focal planes, the temperature of detection port $i$ and lossy port $\LL i$ is typically $< 0.5$ K.
Therefore, when evaluating Eq.~(\ref{equ:mutual_intensity_vs_n}), the occupation number $n(\nu, T)$ due to thermal emission from these detection ports are small at (sub)millimeter frequencies and can be neglected in further calculations. We can then write Eq.~(\ref{equ:mutual_intensity_vs_n}) as
\begin{equation}
    B_{ij}(\nu) = \sqrt{ \eta_i(\nu) \, \eta_j(\nu) } \sum_{k} S^{(i)*}_{k} (\nu) \, S^{(j)}_{k} (\nu) \, n(T_k,\nu)
    \label{equ:mutual_intensity_optical}
\end{equation}
since
\begin{equation}
    S_{ik}(\nu) = S^{\dd i}_{i,\CC i}(\nu) \, S^{(i)}_{k}(\nu) = \sqrt{\eta_i(\nu)} \, S^{(i)}_{k}(\nu) \:.
\end{equation}
Note that we can ignore the overall complex phase of $S^{\dd i}_{i,\CC i}(\nu) = \sqrt{\eta_i(\nu)}$ without loss of generality.

We can define the occupation number at the detector input as an average of the
occupation numbers of all photon sources
\begin{equation}
    n(T_{(i)}, \nu) \equiv \sum_{k} |S^{(i)}_{k}(\nu)|^2 \: n(T_k, \nu) \:,
    \label{equ:effective_brightness}
\end{equation}
since $S^{(i)}_{k}(\nu)$ satisfies the normalization
\begin{equation}
    \sum_{k} |S^{(i)}_{k}(\nu)|^2 = 1
    \: .
\end{equation}
In Eq.~(\ref{equ:effective_brightness}), we define $T_{(i)}$ as the effective brightness temperature of the photons impinging on detector $i$.  This quantity $T_{(i)}$ is generally frequency dependent, and thus
$n(T_{(i)}, \nu)$ does not follow a blackbody spectrum.  However, the photon statistics at each frequency $\nu$ do follow the mixed-state thermal density matrix of temperature $T_{(i)}$ (see Appendix~\ref{app:thermal_density_matrix}), providing the physical foundation for the outcome presented below.

Given the detector quantum efficiency and the input effective brightness temperature, the covariance between detectors $i$ and $j$ in Eq.~(\ref{equ:circuit_shot_wave}) can now be written as
\begin{widetext}
\begin{equation}
\begin{split}
    \sigma_{ij, \mathrm{shot}}^{2}  & = \frac{1}{\tau} \int_{\nu_{1}}^{\nu_{2}} \dd \nu \, (h \nu)^2 \, \eta_{i}(\nu) \, n(T_{(i)}, \nu) \, \delta_{ij} \:, \\
    \sigma_{ij, \mathrm{wave}}^{2}  & = \frac{1}{\tau} \int_{\nu_{1}}^{\nu_{2}} \dd \nu \, (h \nu)^{2}  \, \gamma^{(2)}_{ij}(\nu) \, \eta_{i}(\nu) \, \eta_{j}(\nu) \, n(T_{(i)}, \nu) \, n(T_{(j)}, \nu) \,.
\end{split}
    \label{eq:shot_wave_noise}
\end{equation}
\end{widetext}
As shown in Eq.~(\ref{eq:shot_wave_noise}), the problem of calculating wave noise correlations is reduced to finding the HBT coefficients $\gamma^{(2)}_{ij}(\nu) = | \gamma_{ij}(\nu) |^{2}$, the input mode's effective brightness temperature $T_{(i)}$, and the detector's quantum efficiency $\eta_i(\nu)$.
As shown in Eqs.~(\ref{equ:gamma2_in_circuit_model}) and (\ref{equ:mutual_intensity_optical}), 
the HBT coefficient $\gamma^{(2)}_{ij}(\nu)$ is solely determined by $S^{(i)}_{k}$ and $S^{(j)}_{k}$,
which are in turn defined by the telescope's opticcal configuration and the detector's angular response function.

\begin{figure}[btp]
    \centering
    \includegraphics[width=0.45\textwidth, trim=1cm 3cm 1cm 2cm, clip]{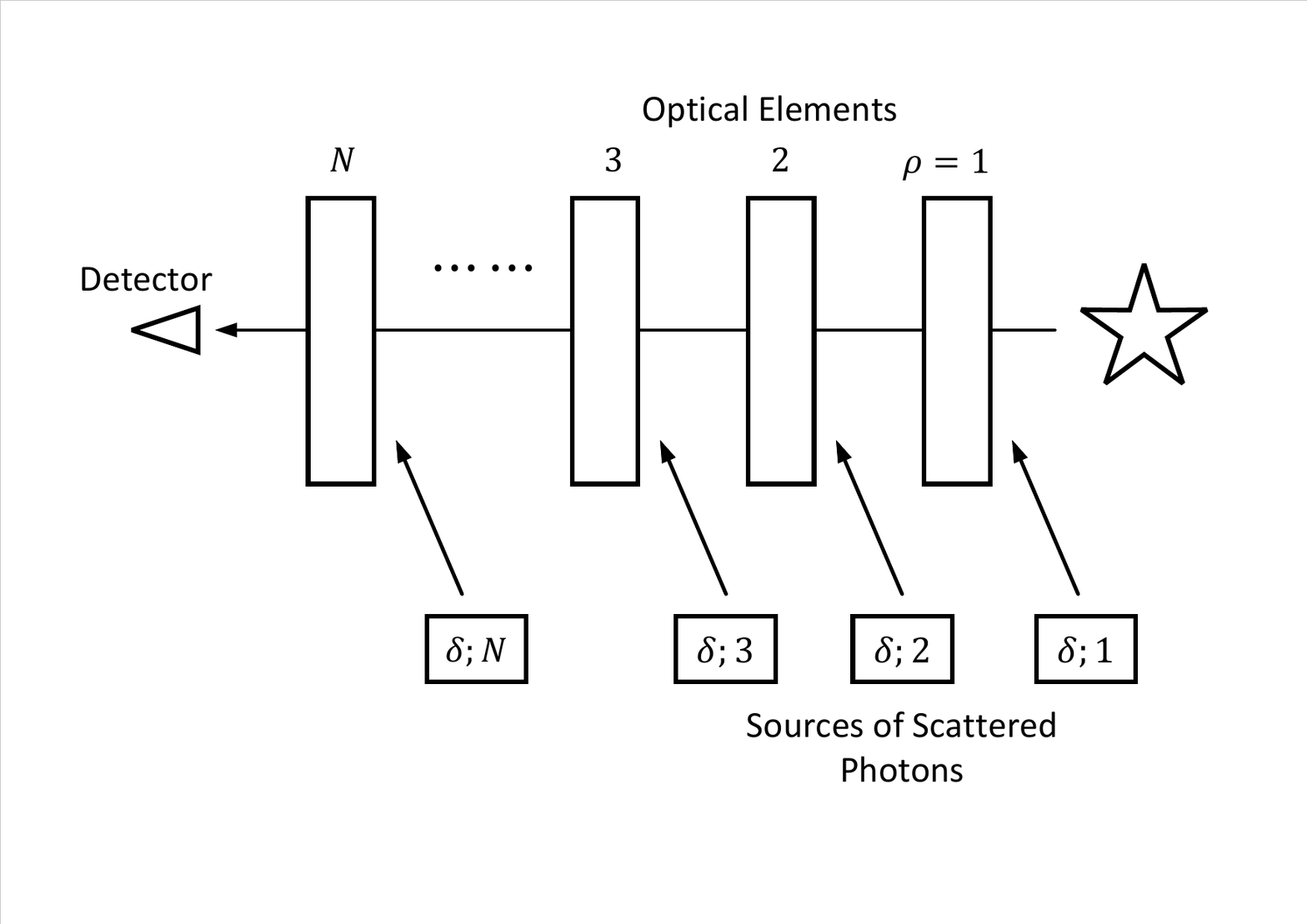}
    \caption{A schematic showing how optical elements contribute to the impinging photons on a detector.
    Each optical element $\rho=1,2,3, \cdots$ has a thermal emissivity $\epsilon_\rho(\nu)$ and temperature $T_\rho$.
    At the same time, each optical element $\rho=1, 2, 3, \cdots$ scatters a fraction $\delta_\rho(\nu)$ of photons from sources
    $\rho=\delta ;\! 1, \, \delta ;\! 2 , \, \delta ;\! 3 , \cdots$ into the line of sight.
    The transmission efficiency of the optical element $\rho$ is then given as $\eta_\rho(\nu) = 1 - \epsilon_\rho(\nu) - \delta_\rho(\nu)$.}
    \label{fig:simple_optical_stack}
\end{figure}

When $i=j$, Eq.~(\ref{eq:shot_wave_noise}) is consistent with the standard bolometer noise model (see, e.g., Ref.~\cite{hill_bolocalc_2018} and references therein). Each optical element can be simply expressed by its transmission, emission, and scattering, as shown in Fig.~\ref{fig:simple_optical_stack}.  The scattering matrix can then be written as
\begin{equation}
    S^{(i)}_{k}(\nu) = 
    \begin{dcases}
    f_k \, \sqrt{ \epsilon_\rho(\nu) } \sqrt{ \mathcal{H}_\rho (\nu) } 
    \: & 
    (k \in \mathcal{M}_\rho)  \\
         f_k \, \sqrt{ \delta_\rho(\nu) } \sqrt{ \mathcal{H}_\rho (\nu) } 
    \: & 
    (k \in \mathcal{M}_{\delta ; \rho}) 
    \end{dcases}
    \label{equ:s_matrix_with_efficiency}
\end{equation}
with
\begin{equation}
    \mathcal{H}_\rho (\nu) \equiv \prod_{l=\rho+1, \cdots, N} \eta_l(\nu) \: ,
\end{equation}
where $\eta_\rho(\nu)$, $\epsilon_\rho(\nu)$, and $\delta_\rho(\nu)$ are the fraction of transmission, emission, and scattering, respectively, of each optical element $\rho$ and satisfy $\eta_\rho(\nu) + \epsilon_\rho(\nu) + \delta_\rho(\nu) = 1$;
$\mathcal{M}_\rho$ and $\mathcal{M}_{\delta ;\rho}$
represent the modes due to thermal emission from element $\rho$ and scattering from element $\rho$ (denoted as $\delta ; \! \rho$), respectively; and $f_k$ is the fractional contribution of each mode given the following normalization
\begin{equation}
    \sum_{k \in \mathcal{M}_\rho} |f_k|^2
    = \sum_{k \in \mathcal{M}_{\delta; \rho}} |f_k|^2
    = 1 \:.
    \label{equ:fractional_mode_normalization}
\end{equation}
By plugging Eqs.~(\ref{equ:s_matrix_with_efficiency})--(\ref{equ:fractional_mode_normalization}) into Eq.~(\ref{equ:effective_brightness}), we obtain\footnote{While the right-hand side of Eq.~(\ref{equ:mutual_itensity_via_eff_scatter})
does not have an explicit dependence on the pixel index $i$, an implicit dependence enters into $\eta_\rho(\nu)$, $\epsilon_\rho(\nu)$, and $\delta_\rho(\nu)$
since different pixels have slightly different viewing angles of and path length differences to each optical element. These differences are minor, however, and can be ignored in most of practical cases.}
\begin{equation}
\begin{split}
    n( & T_{(i)} , \nu) = \\
    & \sum_{\rho = 1, \cdots, N} \mathcal{H}_\rho (\nu) \,
    \left\{\epsilon_\rho (\nu) \, n(T_\rho, \nu) + \delta_\rho (\nu) \, n(T_{\delta; \rho} , \nu ) \right\} .
\end{split}
    \label{equ:mutual_itensity_via_eff_scatter}
\end{equation}

Equations~(\ref{eq:shot_wave_noise}) and (\ref{equ:mutual_itensity_via_eff_scatter}) lead to a formalism consistent with that presented in literature, (e.g., Ref.~\cite{hill_bolocalc_2018}). In our convention defined by Eq.~(\ref{eq:shot_wave_noise}), $\sqrt{\tau} \sigma_{ii}$ is the photon-noise NEP with an S.I. unit of $\mathrm{W\cdot \sqrt{s}}$ and may differ by a factor $\sqrt{2}$ when compared to literature where the NEP is often presented in $\mathrm{W / \sqrt{Hz}}$.


\section{Model optical system}
\label{sec:optical_model}

Using the photon noise formulation in Eq.~(\ref{eq:shot_wave_noise}), we now move to quantify the impact of HBT correlations on the sensitivity of telescopes for mm-wave astronomy. The topic of intensity correlations from astronomical sources with various states of coherence is discussed extensively in the literature~\cite{baltes_spectral_1976,carter_coherence_1975,carter_coherence_1977,wolf_angular_1975, wolf_radiometric_1976, wolf_coherence_1978,agarwal_coherence_2004}, and in the sections that follow, we apply these findings to mm-wave telescope design. Modern mm-wave telescopes employ a wide variety of lens and mirror systems, infrared filter stacks, anti-reflection coatings, and sensing architectures. Despite this variety in real experiments, we can distill a few key instrument characteristics to create a simple yet representative optical system in which to study the sensitivity impact of HBT correlations. More specifically, this simple system, characterized by its focal ratio and the effective brightness temperatures inside and outside of its pupil stop, emulates the radiation environment at the focal plane of general, modern millimeter and submillimeter telescopes with sufficient accuracy.

Our goal is to calculate intensity correlations within a practical telescope system with $\sim 10\,\%$ accuracy. As shown in the following sections, the HBT correlation coefficient has a maximum sensitivity impact of $\mathcal{O}(10\,\%)$; therefore our HBT accuracy goal maps to a sensitivity accuracy of $\sim 1\,\%$, which is typically sufficient for the purposes of telescope design.
While our idealized optics and detector focal planes may not exactly reproduce those of real telescopes, they do capture the characteristics needed to calculate the HBT coefficient of real systems with $\sim 10\,\%$ accuracy.

\begin{figure}[btp]
    \centering
    \includegraphics[width=0.4\textwidth, trim=7cm 4cm 7cm 4cm, clip]{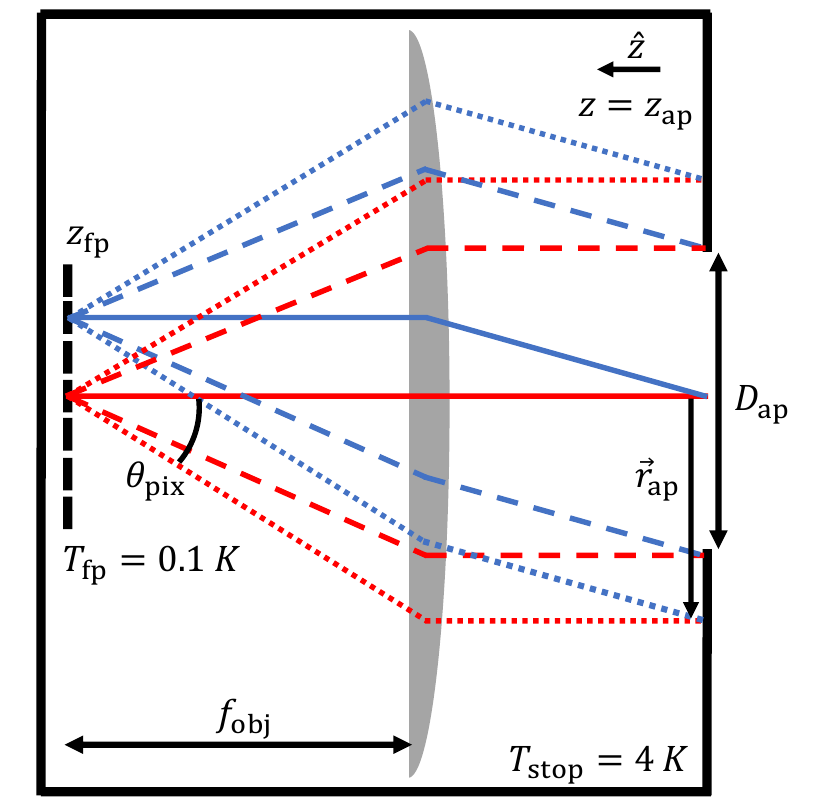}
    \caption{The assumed optical model for all calculations and simulations in this paper. The model includes an objective lens, aperture stop, and focal plane filled with an array of sensing antennas coupled to planar detectors. In the reverse-time sense, each pixel emits a collection of rays, defined by spherical coordinate $(\theta_{\mathrm{pix}}, \phi_{\mathrm{pix}})$, which the objective uniquely maps onto aperture-plane coordinate $\vec{r}_{\mathrm{a}}$ with idealized polarization fidelity and telecentricity. The system is enclosed in a black box of temperature $T_{\mathrm{stop}}$, and the focal plane is cooled to $T_{\mathrm{fp}}$.}
    \label{fig:simple_optics}
\end{figure}


\subsection{Telescope}
\label{sec:telescope_model}

The assumed example telescope model is depicted in Fig.~\ref{fig:simple_optics}. It consists of an objective lens with focal length $f_{\mathrm{obj}}$ within a blackened enclosure at $T_{\mathrm{stop}} = 4$~K. The cold box has a circular aperture of diameter $D_{\mathrm{ap}}$ at $z = z_{\mathrm{ap}}$ that truncates incoming radiation from external field-filling sources. The objective lens, which is both cold and transparent, focuses the aperture-truncated radiation onto a circular focal plane at $z = z_{\mathrm{fp}}$ with a size determined by the telescope's plate scale. The focal plane houses an array of close-packed detector pixels with diameter $D_{\mathrm{pix}}$ operating at $T_{\mathrm{fp}} = 0.1$~K. The detector + telescope system is assumed to be diffraction limited such that the optical throughput per mode is $A \Omega = \lambda^{2}$.

This model optical system does not include many common features of real telescopes---such as fore-optics, thermal filters, or additional lenses---which are needed to form high-fidelity images over a moderate field of view (FOV). Such details are experiment-dependent and are therefore beyond the scope of this paper, but we can capture their effects by imposing several assumptions onto our simple system. These assumptions are not strictly necessary to calculate photon-noise correlations, as the scattering matrix formalism in Sec.~\ref{sec:zmu_circuit} is completely general, but they simplify the correlation calculation significantly while encapsulating the salient features of practical instruments.

Firstly, we assume that all sources---both external and internal to the telescope---are isothermal blackbody emitters large enough to uniformly illuminate the aperture across the telescope's FOV. The assumption of blackbodies allow us to readily evaluate each mode's occupation number using the Bose-Einstein distribution $n(\nu, T_{\mathrm{b}})$ in Eq.~(\ref{eq:bose_einstein}) given each source's effective brightness temperature $T_{\mathrm{b}}$. The assumption that each source is FOV-filling\footnote{An obvious exception is the aperture stop, which we treat separately from aperture-filling radiation.} and isothermal generalizes the correlation integrals that follow and is a good approximation for experiments whose incoming photon power is mainly from extended sources. In practice, thermal gradients develop across optical elements and atmospheric brightness varies with elevation and cloud structure; however, these variations are typically small and experiment-dependent and are therefore beyond the scope of this paper. 

Secondly, we assume a diffraction-limited, single-moded optical system that converts stop-truncated plane waves into spherical waves converging onto a telecentric focal plane. In other words, pixel rays with angle $(\theta_{\mathrm{pix}}, \phi_{\mathrm{pix}})$ are mapped via the objective lens onto parallel rays with aperture-plane location $\vec{r}_{\mathrm{ap}}$ (see Fig.~\ref{fig:simple_optics}). In this configuration, the optical path length between any given detector pixel and a spot on the aperture stop is identical regardless of $(\theta_{\mathrm{pix}}, \phi_{\mathrm{pix}})$ or $\vec{r}_{\mathrm{ap}}$, which simplifies the calculations that follow. We note that when re-imaging optics and a pupil stop are employed, modes which map onto spherical waves at the focal plane do not in general correspond to plane waves passing through the pupil stop. However, such reimaging optics can always be modeled as a simplified equivalent system with an aperture stop provided that each optic's clear-aperture diameter is large enough to pass all pupil-permitted modes.

Thirdly, we assume an ideal aperture stop, such that all detector pixels have the same mapping between ray angle $(\theta_{\mathrm{pix}}, \phi_{\mathrm{pix}})$ and aperture plane location $\vec{r}_{\mathrm{ap}}$ (see Fig.~\ref{fig:simple_optics}). In other words, the aperture illumination is identical regardless the detector pixel location.\footnote{In theory, the necessary condition for an ideal aperture is for the mapping to be identical only on the stop circumference, but in practice, when this condition is met, the mapping becomes identical within the aperture stop as well.} Strictly speaking, this condition is not generally satisfied for a system with a large FOV, as telecentricity and aperture truncation may differ significantly between the central and peripheral regions of the focal plane. However, as we discuss later, photon-noise correlations arise predominantly between neighboring pixels where such non-idealities are negligible.

\begin{figure}[tbp]
    \centering
    \includegraphics[width=0.45\textwidth, trim=5cm 4cm 3cm 4cm, clip]{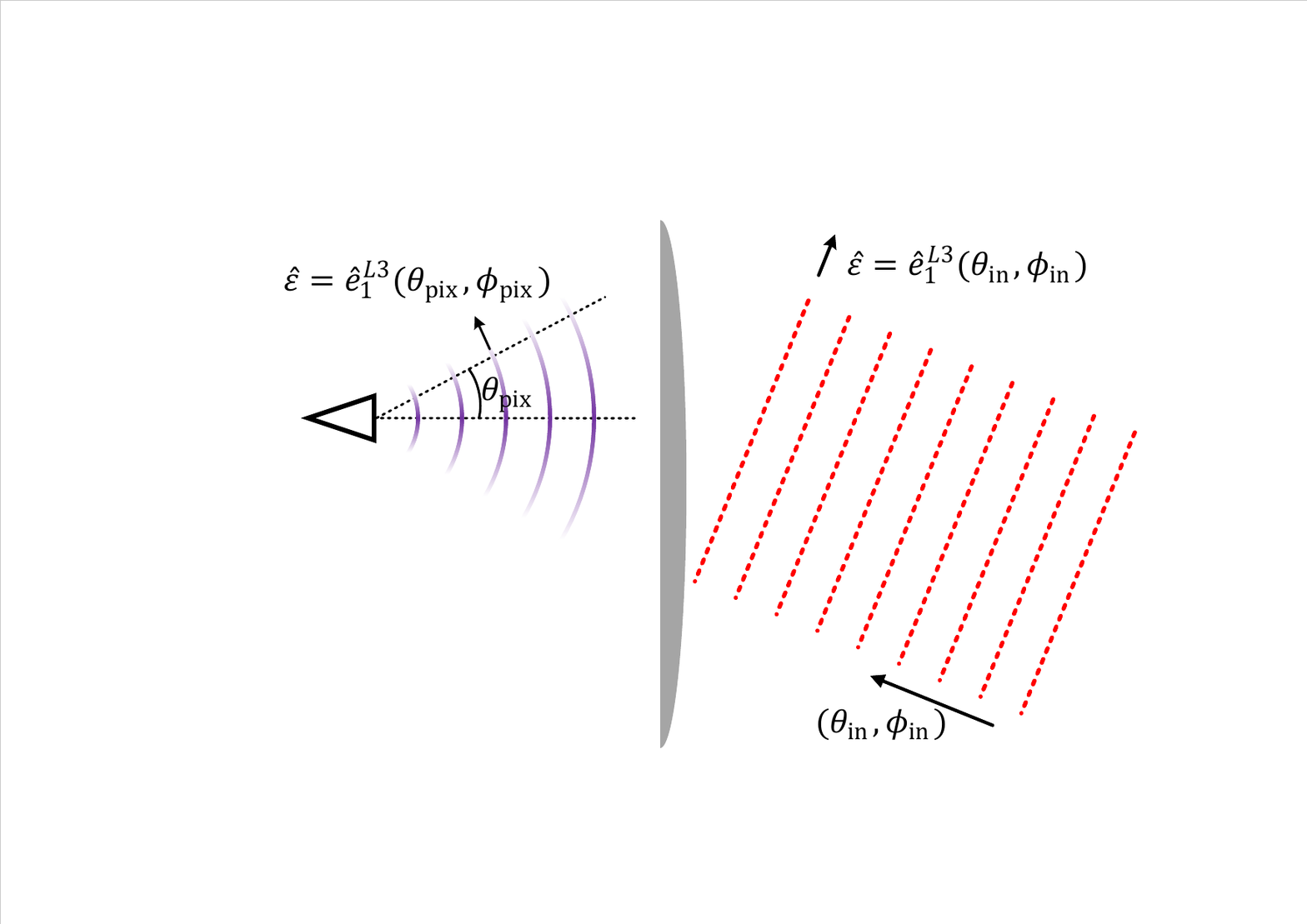}
    \caption{A schematic of the ideal optics that focus incident plane waves onto detectors as spherical waves with Ludwig-3 polarization distributions. In this example, the incident wave has a propagation direction $(\theta_{\rm in}, \phi_{\rm in})$ and polarization direction $\hat{\varepsilon} = \hat{e}_1^{L3}(\theta_{\rm in}, \phi_{\rm in})$,
    and it is converted to a spherical wave with polarization direction  
    $\hat{\varepsilon} = \hat{e}_1^{L3}(\theta_{\rm pix}, \phi_{\rm pix})$
    for all $(\theta_{\rm pix}, \phi_{\rm pix})$
    mapped within the aperture stop.}
    \label{fig:ideal_optics}
\end{figure}

Fourthly, we assume that the telescope optics achieve polarization fidelity across the focal plane.
As shown in Fig.~\ref{fig:ideal_optics}, incident linearly-polarized plane waves with propagation direction $(\theta_{\rm in}, \phi_{\rm in})$ and orthogonal polarization vectors $\hat{e}_1^{L3}(\theta_{\rm in}, \phi_{\rm in})$ and $\hat{e}_2^{L3}(\theta_{\rm in}, \phi_{\rm in})$ are focused onto detector pixels as spherical waves with Ludwig-3 polarization distributions~(Eq.~\ref{eq:ludwig_polariztion}~\cite{ludwig_definition_1973}):
\begin{equation}
\begin{split}
    \hat{e}_1^{L3}(\theta_{\rm in}, \phi_{\rm in}) & \rightarrow \hat{e}_1^{L3}(\theta_{\rm pix}, \phi_{\rm pix}) \: , \\
    \hat{e}_2^{L3}(\theta_{\rm in}, \phi_{\rm in}) & \rightarrow \hat{e}_2^{L3}(\theta_{\rm pix}, \phi_{\rm pix}) \: .
\end{split}
\label{eq:ludwig3_spherical_optics}
\end{equation}
For an on-axis incident plane-wave, this relation simplifies to
\begin{equation}
    \hat{x} \rightarrow \hat{e}_1^{L3}(\theta_{\rm pix}, \phi_{\rm pix}) \: , \quad  
    \hat{y} \rightarrow \hat{e}_2^{L3}(\theta_{\rm pix}, \phi_{\rm pix}) \: .
\end{equation}
It follows from this assumption that HBT correlations cannot develop between orthogonal polarimeters.
In practice, some cross polarization does exist within real telescopes, and the degree of polarization leakage can vary across the focal plane. However, modern polarimetry experiments are specifically designed to suppress cross polarization~\cite{Mizuguchi2005OffsetAntenna,Dragone1978OffsetDiscrimination,tran_comparison_2008}, especially over localized areas on the focal plane where intensity correlations are important.

Finally, we assume that the objective lens is cold and transparent such that its emission
and scattering terms
are negligible compared to those of other internal and external thermal sources.


\subsection{Focal plane}
\label{sec:focal_plane_model}

\begin{figure}[tbp]
    \centering
    \includegraphics[width=0.45\textwidth, trim=3.0cm 1.7cm 0cm 3.2cm, clip]{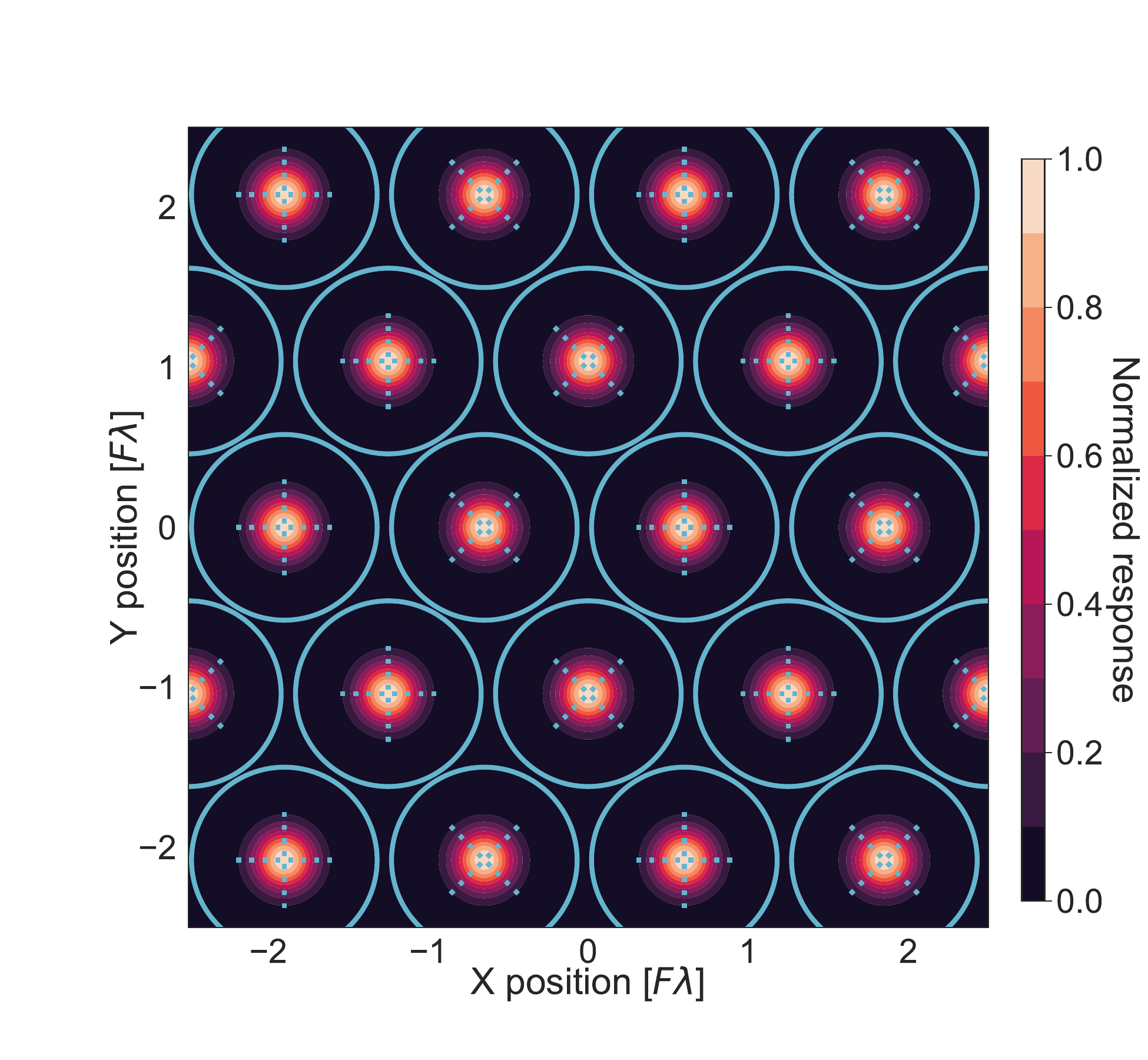}
    \caption{The assumed layout of detector pixels on the focal plane at $D_{\mathrm{pix}} = 1.2 F \lambda$ spacing. Each pixel's angular response is assumed to be a Gaussian whose width scales with pixel diameter, as described in Eq.~(\ref{eq:detector_gaussian_beam}). Additionally, each pixel has two polarimeters (dotted lines) that sense orthogonal polarizations and whose noise outputs do not correlate, given the idealized optical system described in Sec.~\ref{sec:optical_model}. Neighboring pixels are rotated by $\pm$ 45 deg to minimize Stokes Q HBT correlations.}
    \label{fig:focal_plane_layout}
\end{figure}

The assumed focal plane model is shown in Fig.~\ref{fig:focal_plane_layout}. We assume single-moded, dual-polarization detector pixels with circular apertures and diffraction-limited Gaussian beams. Each pixel's angular response function is determined solely by its beam waist $w_{0}$ and takes the far-field form
\begin{equation}
    E(\theta) \approx E_{0} \; \mathrm{exp}\left[ -\frac{\theta^{2}}{(\lambda / \pi w_{0})^{2}} \right] \, .
    \label{eq:detector_gaussian_beam}
\end{equation}
Each polarimeter has a diffraction-limited throughput of $A \Omega = \lambda^{2}$, regardless of the pixel's aperture size, and we assume that the beam pattern in Eq.~(\ref{eq:detector_gaussian_beam}) is symmetric between the antenna's $E$ and $H$ planes and follows the Ludwig-3 polarization response~(Eqs.~\ref{equ:ideal_detector} and \ref{equ:ideal_detector2}).\footnote{A Ludwig-3 polarization beam pattern often results from an angular response function with $E$-plane/$H$-plane symmetry. See also the footnote in Sec.~\ref{sec:simple_with_pol}.}  A larger/smaller pixel results in a narrower/wider far-field response, and we linearly relate the pixel diameter $D_{\mathrm{pix}}$ to the beam waist via a scaling constant $w_{f}$
\begin{equation}
    w_{0} = \frac{D_{\mathrm{pix}}}{w_{f}}
    \label{eq:wf} \, .
\end{equation}
Typical mm-wave detector pixels, such a corrugated feedhorns, spline-profiled feedhorns, and lenslet-coupled planar antennas achieve $w_{f} \approx 3$, which we assume for the calculations that follow.

Plugging Eq.~(\ref{eq:wf}) into Eq.~(\ref{eq:detector_gaussian_beam}) yields a simple relationship between $D_{\mathrm{pix}}$ and aperture stop spillover efficiency
\begin{equation}
    \eta_{\mathrm{ap}} = \frac{\int_{0}^{\theta_{\mathrm{stop}}} E^{2}(\theta) \, \dd \theta}{\int_{0}^{\pi/2} E^{2}(\theta) \, \dd \theta} = 1 - \exp \left[ -\frac{\pi^{2}}{2} \left( \frac{D_{\mathrm{pix}}}{F \lambda w_{\mathrm{f}}} \right)^{2} \right] \, ,
    \label{eq:aperture_spill_efficiency}
\end{equation}
\noindent
where $F \equiv D_{\mathrm{ap}} / f_{\mathrm{obj}}$ is the F-number at the focal plane and $\theta_{\mathrm{stop}} = \arctan \left[ 1 / (2 F) \right]$. This assumption of Gaussian spillover efficiency does not necessarily hold when $D_{\mathrm{pix}} < \lambda$, as diffraction at the pixel edges will create substantial ringing in the far-field beam pattern. However, in an effort to remain agnostic to the specifics of the detector coupling architecture, we assume that Eqs.~(\ref{eq:detector_gaussian_beam}) and~(\ref{eq:aperture_spill_efficiency}) remain valid for all values of $D_{\mathrm{pix}}$ in the calculations that follow.

While detector pixels can be arranged in a variety of ways, we must select a specific focal plane arrangement to find $\left| \gamma_{ij}(\nu) \right|^{2}$ explicitly. In the calculations that follow, we assume hex-packed circular pixels, and we assume that pixel pitch is equal to pixel diameter $p_{\mathrm{pix}} = D_{\mathrm{pix}}$. This assumption allows us to relate pixel packing density to pixel size as $n_{\mathrm{pix}} \propto D_{\mathrm{pix}}^{-2}$, where $n_{\mathrm{pix}}$ is the number of pixels per unit focal plane area. In practice, a small amount of dead space typically exists between pixels that does not scale with pixel size, and in this case, a more complex relationship between $n_{\mathrm{pix}}$ and $D_{\mathrm{pix}}$ is needed~\cite{padin_mapping_2010}. However, these details are experiment specific and are therefore beyond the scope of the following discussions.

We note that the correlation calculation in Sec.~\ref{sec:correlations} does not rely on most of the assumptions presented in this subsection. Specifically, only the detector beam's assumed polarization properties are relevant to the HBT coefficient estimation. All the other assumptions related to the shape of the detector beam, the beam waist,  and pixel packing are used only to calculate an explicit instrument sensitivity in Sec.~\ref{sec:sensitivity} and Sec.~\ref{sec:implications_for_experiment_design}.

%
\begin{figure}[t!]
    \centering
    \includegraphics[width=0.45\textwidth, trim=7cm 4cm 5cm 4cm, clip]{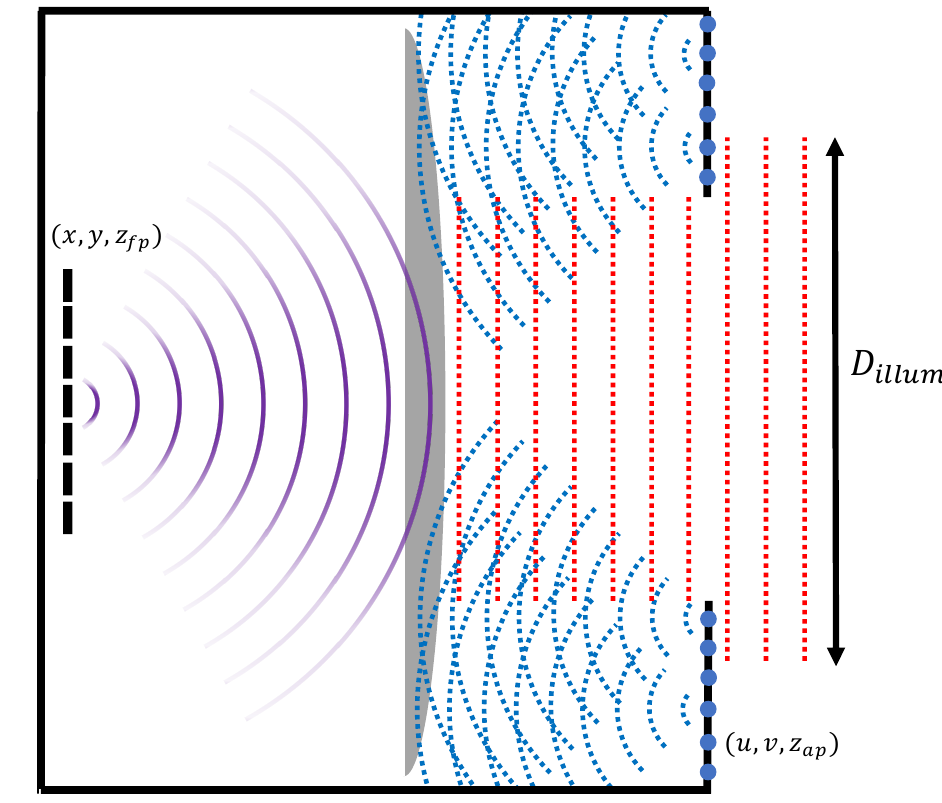}
    \caption{A schematic of the aperture, stop, and pixel radiation models. Radiation within the aperture is decomposed into a basis set of plane waves, and the normally incident mode $(k_{x}, k_{y}, k_{z}) = (0, 0, 2\pi/\lambda)$ is depicted here. The plane waves are assumed to uniformly illuminate a diameter $D_{\mathrm{illum}} > D_{\mathrm{ap}}$, and each mode's Gaussian-random amplitude is defined by its effective brightness temperature $T_{\rm ap}$. Radiation from the stop is generated by a collection of blackbody point sources, which emit spherical Lambertian wavelets located at $(x,y,z) = (u, v, z_{ap})$ with $\sqrt{u^{2} + v^{2}} \geq D_{\mathrm{ap}} / 2$. Both the aperture's plane-wave modes and the stop's wavelet modes couple to the objective lens, which focuses the thermal light onto the focal plane. A detector pixel then senses the incoming spherical wave via the Gaussian angular response in Eq.~(\ref{eq:detector_gaussian_beam}).}
    \label{fig:aperture_stop_radiation}
\end{figure}

\section{Correlation calculation}
\label{sec:correlations}
Given the optical and detector models presented in Sec.~\ref{sec:optical_model}, we now find the correlation patterns at the focal plane due to thermal radiation within the aperture and from the stop. A schematic of the radiation model for the central detector pixel is shown in Fig.~\ref{fig:aperture_stop_radiation}.


\subsection{Stop radiation}
\label{sec:stop_radiation}

The stop is located in the far field of the detectors and is effectively a black, annular source with temperature $T_{\mathrm{stop}}$. We therefore model the stop as a collection of infinitesimal, Gaussian-random, uncorrelated thermal emitters that generate Lambertian spherical wavelets, as shown in Fig.~\ref{fig:aperture_stop_radiation}. These point sources represent atomic thermal motion within the stop's absorbing material, and their wavelets superpose to form incoherent waves that the objective lens focuses onto the focal plane.

The above stop radiation treatment relies on two assumptions, which we justify here. First, while we consider the stop as a collection of uncorrelated thermal sources, it is known that blackbody radiators have non-zero correlation over the distance of a wavelength~\cite{mehta_coherence_1964,carter_coherence_1975,baltes_spectral_1976,steinle_radiant_1977}.
However, as we will show in Sec.~\ref{sec:aperture_radiation}, Sec.~\ref{sec:corr_patterns}, and Appendix~\ref{app:partial_coherence}, this discrepancy leads to negligible errors for the calculations in this paper.
Second, we assume for simplicity that all stop radiation reaches the detectors by propagating through all optics between the aperture plane and the focal plane, which impractically requires infinite optical throughput. That said, the cold box's radiation environment has temperature $T_{\rm stop}$, and the detectors sense that radiation when $\theta_{\mathrm{pix}} > \theta_{\mathrm{stop}}$, regardless of the optical configuration. Therefore, our simplifying assumption of infinite optical throughput accurately accounts for stop-generated photons at each detector's input.



\subsection{Aperture radiation}
\label{sec:aperture_radiation}
Radiation incident on the sky side of the aperture (right side of Fig.~\ref{fig:aperture_stop_radiation}) can be regarded as blackbody emission\footnote{We assume that the
incident radiation is largely unpolarized,
or $I \gg \sqrt{Q^2 + U^2}$ in terms of Stokes $I, Q, U$.  This is a good approximation for millimeter and submillimeter telescopes.  Consideration
of polarized sources, which is outside the scope of this paper, can be found elsewhere~\cite{tervo_van_2013}.
} with an effective brightness temperature $T_{\rm (ap)}$.
Here, $T_{\rm (ap)}$ is defined such that $n(T_{\rm (ap)}, \nu)$ is the mean occupation number of the blackbody radiation within the aperture. This mean occupation number can be written by adopting the definitions in Sec.~\ref{sec:photon_noise_correlations} (Fig.~\ref{fig:simple_optical_stack} and Eqs.~\ref{equ:s_matrix_with_efficiency}--\ref{equ:fractional_mode_normalization}) and using $\mathcal{H}_\rho(\nu)$, $\epsilon_\rho(\nu)$, $\delta_\rho(\nu)$ as
\begin{widetext}
\begin{equation}
    n(T_{\rm (ap)}, \nu) = \frac{\sum_{\rho<\rho_{\rm ap}} \left\{\epsilon_\rho (\nu) \, \mathcal{H}_\rho (\nu) \, n(T_\rho, \nu) + \delta_\rho (\nu) \, \mathcal{H}_{\rho} (\nu) \, n(T_{\delta;\rho}, \nu) \right\}}
    {\sum_{\rho<\rho_{\rm ap}} \left\{ \epsilon_\rho (\nu) \, \mathcal{H}_\rho (\nu) + \delta_\rho (\nu) \, \mathcal{H}_{\rho} (\nu) \right\} }
    \: ,
    \label{equ:t_aperture_definition}
\end{equation}
\end{widetext}
where $\rho=\rho_{\rm ap}$ labels the aperture stop element.
This blackbody radiation can be decomposed into plane-wave modes,  
and the FOV-filling nature of the sources (Sec.~\ref{sec:telescope_model}) ensures the above stated radiation property for all plane-wave modes whose propagation direction is within the FOV.\footnote{While  plane-wave modes with propagation vector outside the FOV may have different radiation properties, they are irrelevant in our context of evaluating photon correlation among detectors.}

We then consider a virtual, infinitely large sheet at $z=z_{\rm ap}$ that consists of Gaussian-random, infinitesimal, uncorrelated thermal emitters of temperature $T_{\rm (ap)}$ that generate Lambertian spherical wavelets. Radiation from these virtual emitters has equivalent statistical properties to the sky-side radiation in Fig.~\ref{fig:aperture_stop_radiation} so long as the distances from the sheet are significantly larger than the wavelength $\lambda$.
To calculate this virtual sheet's aperture truncation, we can simply remove elements outside of the aperture area $\sqrt{x^{2} + y^{2}} \geq D_{\mathrm{ap}} / 2$.
In Sec.~\ref{sec:corr_patterns} and Fig.~\ref{fig:hbt_corr_coeff}, we compare aperture-truncated plane waves with this sheet of emitters and demonstrate their equivalence.

There is a small error that arises from the presented aperture emitter treatment. As noted above, the virtual sheet's statistics become that of blackbody radiation at distances sufficiently larger than $\lambda$, and as noted in Sec.~\ref{sec:stop_radiation}, non-zero correlations arise between emitters separated by $\lesssim \lambda$.  Thus, 
the emitter-sheet model deviates from that of aperture truncation in the region of $z\simeq z_{\rm ap}$ and $D_{\mathrm{ap}} / 2 - \lambda \lesssim  \sqrt{x^{2} + y^{2}} \lesssim D_{\mathrm{ap}} / 2 + \lambda$. Therefore when $\lambda \gtrsim D_{\rm ap}$, aperture truncation produces additional non-trivial correlations that include polarization~\cite{chuss_diffraction_2008}, but when $\lambda \ll D_{\mathrm{ap}}$, as is true for any reasonable telescope design,
the correction can be safely neglected.

In summary, the VCZT coefficient $\gamma_{ij}(\nu)$ can be calculated using the radiation field from a sheet of infinitesimal thermal emitters at $z = z_{\rm ap}$ with temperature $T_{\mathrm{stop}}$ at
$\sqrt{x^{2} + y^{2}} \geq D_{\mathrm{ap}} / 2$ and $T_{(\mathrm{ap})}$ at $\sqrt{x^{2} + y^{2}} < D_{\mathrm{ap}} / 2$.


\subsection{Intensity correlation patterns}
\label{sec:corr_patterns}

Here we calculate spatial correlation patterns for the geometry in Fig.~\ref{fig:simple_optics} while neglecting the polarization degree of freedom, similarly to Sec.~\ref{sec:simple_example_wo_pol}. We then consider polarization in the next subsection. Consider the classical-wave electric field amplitude $E_i$ detected by detector $i$ at location $(x,y,z) = (x_i, y_i, z_{\rm fp})$. The detected partial amplitude $\Delta E_i$ due to an infinitesimal source area at the aperture plane $(x,y,z) = (u, v, z_{\rm ap})$ can then be written as
\begin{equation}
    \begin{split}
    \Delta E_i = \, & \, \mathcal{C} \, G_i(u, v) \,
    \sqrt{\cos \theta_{i}}
    \, \\
    & \cdot \, e^{[2\pi i \nu (u \sin \theta_i \cos \phi_i + v \sin \theta_i \sin \phi_i) / c]} \, \Delta u \Delta v
    \:.
    \end{split}
    \label{equ:classical_coupling_in_telescope}
\end{equation}
Here, $G_i(u, v)$ is the aperture illumination function for detector $i$, and $(\theta_i, \phi_i)$ denotes the propagation direction of the incident plane wave focused by the objective lens onto detector $i$, which can be related to the detector's focal plane position $(x_{i}, y_{i})$ via the objective's focal length $f_{\rm obj}$ as
\begin{equation}
    \begin{split}
        \sin \theta_i \cos \phi_i \simeq \frac{x_i}{f_{\rm obj}} & = \frac{x_i}{F D_{\rm ap}} \:,
        \\
        \sin \theta_i \sin \phi_i \simeq \frac{y_i}{f_{\rm obj}} & = \frac{y_i}{F D_{\rm ap}}           \:.
  \end{split}
  \label{equ:thetai_phii_vs_F_xi_yi}
\end{equation}

The illumination function $G_i(u, v)$ is mapped from the detector's angular-response function (Eq.~(\ref{eq:detector_gaussian_beam})) via the telescope's optics.  We adopt the following normalization
\begin{equation}
    \iint \dd u \, \dd v \,  |G_i(u, v)|^2 = 1 \:,
\end{equation}
which leads to
\begin{equation}
    \iint_{\sqrt{u^2 + v^2} < D_{\rm ap}/2}
    \!\!\!\!\!\! \!\!\!\!\!\!
    \dd u \, \dd v \,  |G_i(u, v)|^2 = \eta_{\rm ap} \:,
\end{equation}
where the aperture stop spillover efficiency $\eta_{\mathrm{ap}}$ is defined in Eq.~(\ref{eq:aperture_spill_efficiency}).

The non-polarized mutual intensity between detectors $i$ and $j$ can be written as a sum of contributions from the aperture radiation and stop radiation
\begin{equation}
    B^{\rm np}_{ij}(\nu) = B^{\rm np}_{{\rm ap}, ij}(\nu) + B^{\rm np}_{{\rm stop}, ij}(\nu) \:.
\end{equation}
The aperture-radiation component can then be written as
\begin{widetext}
\begin{equation}
\begin{split}
B^{\rm np}_{{\rm ap}, ij}(\nu)
    & = |\mathcal{C}|^2
    \sqrt{\cos \theta_{i} \cos \theta_{j}} \, n(T_{\rm (ap)}, \nu) 
    \iint_{\sqrt{u^2 + v^2} < D_{\rm ap}/2}
    \!\!\!\!\!\!\!\!\!\!\!\! \dd u \, \dd v \, G_i^*(u, v) \, G_j(u, v) \,
    e^{2\pi i (u x_{ji} + v y_{ji})/D_{\rm ap} F \lambda} \:,
    \\
    & \simeq 
    |\mathcal{C}|^2
    \sqrt{\cos \theta_{i} \cos \theta_{j}} \, n(T_{\rm (ap)}, \nu) 
    \iint_{\sqrt{u^2 + v^2} < D_{\rm ap}/2}
    \!\!\!\!\!\!\!\!\!\!\!\! \dd u \, \dd v \, |G_i(u, v)|^2 \,
    e^{2\pi i \, p_{ij} u /D_{\rm ap} F \lambda} \: ,
\end{split}
\label{equ:mutual_intensity_aperture}
\end{equation}
\end{widetext}
where $x_{ji} \equiv x_j - x_i$, $y_{ji} \equiv y_j - y_i$,
$\lambda \equiv c/\nu$, and $p_{ij} \equiv \sqrt{x_{ji}^2 + y_{ji}^2}$, and where the second equality assumes that
$G_i(u, v) \simeq G_j(u, v)$ and that $G_i(u, v)$ is approximately circularly symmetric. 
The VCZT coefficient
(Eq.~(\ref{equ:gamma1_in_circuit_model})) for the aperture radiation is
\begin{equation}
\begin{split}
    \gamma^{\rm np}_{{\rm ap}, ij}(\nu)
    & =
    \frac{B^{\rm np}_{{\rm ap}, ij}(\nu)}
    {\sqrt{B^{\rm np}_{{\rm ap}, ii}(\nu) \,
    B^{\rm np}_{{\rm ap}, ii}(\nu)}}
    \:,
    \\
    & = \frac{1}{\eta_{\rm ap}}
    \iint_{\sqrt{u^2 + v^2} < D_{\rm ap}/2}
    \!\!\!\!\!\!\!\!\!\!\!\!\!\!\!\!\!\!\!\!\!\!\!\!\!\! \dd u \, \dd v \; |G_i(u, v)|^2 \,
    e^{2\pi i \, p_{ij} u /D_{\rm ap} F \lambda} \:.
\end{split}
\label{eq:aperture_coherence}
\end{equation}
For our purpose of estimating the degree of coherence, a Gaussian illumination function is a good enough approximation:
\begin{equation}
    G_i(u, v) = \frac{1}{\sqrt{\pi \sigma_{\rm ap}^2}}
    \exp \left( - \frac{u^2 + v^2}{2 \sigma_{\rm ap}^2} \right) \: .
    \label{eq:gaussian_illum_func}
\end{equation}
In an extreme example of a flat illumination function $G_i(u, v) = \mathrm{const.}$ (or $\sigma_{\rm ap} \rightarrow \infty$),
Eq.~(\ref{eq:aperture_coherence}) reduces to 
\begin{equation}
  \gamma^{\rm np}_{{\rm ap}, ij}(\nu)
  = \frac{2 J_1 \left(\pi p_{ij} / F \lambda \right)}{\pi p_{ij} / F \lambda}
    \:,
    \label{equ:flat_illum_approx}
\end{equation}
where $J_1(x)$ is the Bessel function of the first kind. This form is equivalent to the intensity diffraction pattern from a circular aperture, mirroring the well-known correspondence between diffraction and coherence formalisms (e.g., see Ref.~\cite{wolf_2007}).
As discussed in Appendix~\ref{app:illumination_flat_vs_realistic}, Eq.~(\ref{equ:flat_illum_approx}) turns out to be a very good approximation for a general Gaussian illumination function $G_i(u, v)$ when we assume the relation between the detector's beam and pixel diameter discussed in Sec.~\ref{sec:optical_model}. We thus use this handy approximation hereafter.

The stop-radiation component $B^{\rm np}_{{\rm stop}, ij}(\nu)$ and $\gamma^{\rm np}_{{\rm stop}, ij}(\nu)$
can be calculated similarly to Eqs.~(\ref{equ:mutual_intensity_aperture}) and (\ref{eq:aperture_coherence}) but instead using $T_{\rm stop}$ and integrating over $\sqrt{u^2 + v^2} \geq  D_{\rm ap}/2$:
\begin{equation}
\begin{split}
    & \gamma^{\rm np}_{{\rm stop}, ij}(\nu)
    \\
    & \: = (1-\eta_{\rm ap})^{-1}
    \iint_{D_{\rm ap}/2 \leq \sqrt{u^2 + v^2}}
    \!\!\!\!\!\!\!\!\!\!\!\!\!\!\!\!\!\!\!\!\!\!\!\!\!\! \dd u \, \dd v \; |G_i(u, v)|^2 \,
    e^{2\pi i \, p_{ij} u /D_{\rm ap} F \lambda}
\end{split}
\label{equ:stop_radiation_coherence}
\end{equation}
and
\begin{equation}
\begin{split}
    B^{\rm np}_{{\rm stop}, ij}(\nu)
    = & |\mathcal{C}|^2
    \sqrt{\cos \theta_{i} \cos \theta_{j}} \, n(T_{\rm stop}, \nu)
    \\
    & \quad
    \cdot (1-\eta_{\rm ap}) \; \gamma^{\rm np}_{{\rm stop}, ij}(\nu) \:.    
\end{split}
\end{equation}
As discussed in Appendix~\ref{app:illumination_flat_vs_realistic}, this term can also be approximated using a flat illumination and limiting the range of integration to $D_{\rm ap}/2 \leq \sqrt{u^2 + v^2} \leq D_{\rm ap}/2 + \sigma_{\rm ap}$.  We then obtain the approximation
\begin{equation}
    \begin{split}
    & \gamma^{\rm np}_{{\rm stop}, ij} (\nu)
    \\
    & \; \simeq \frac{F^2 F'^2}{F^2 - F'^2}
    \left\{ \frac{1}{F'^2}\frac{2 J_1 (\pi p_{ij} / F' \lambda)}{\pi p_{ij} / F' \lambda}
    -  \frac{1}{F^2}\frac{2 J_1 (\pi p_{ij} / F \lambda)}{\pi p_{ij} / F \lambda} \right\}
    \end{split}
    \label{eq:stop_coherence_approx}
\end{equation}
with
\[
   F' \equiv \frac{D_{\rm ap}}{D_{\rm ap} + 2 \sigma_{\rm ap}} F \:.
\]

\begin{figure*}[tbp]
    \begin{subfigure}
        \centering
        \includegraphics[width=0.45\textwidth, trim=0.3cm 1.4cm 3cm 3.5cm, clip]{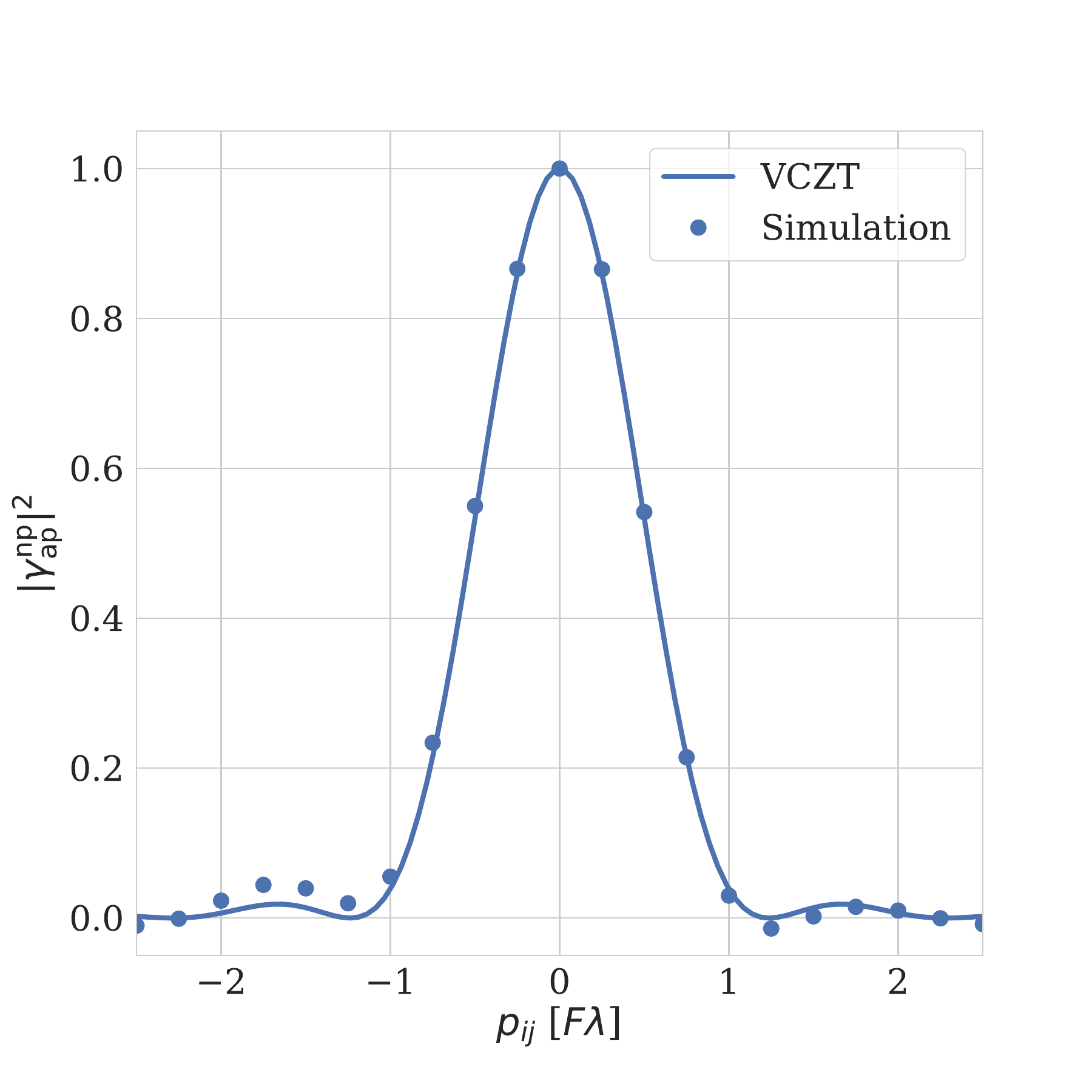}
    \end{subfigure}
    \hspace{0.03\textwidth}
    \begin{subfigure}
        \centering
        \includegraphics[width=0.45\textwidth, trim=0.3cm 1.4cm 3cm 3.5cm, clip]{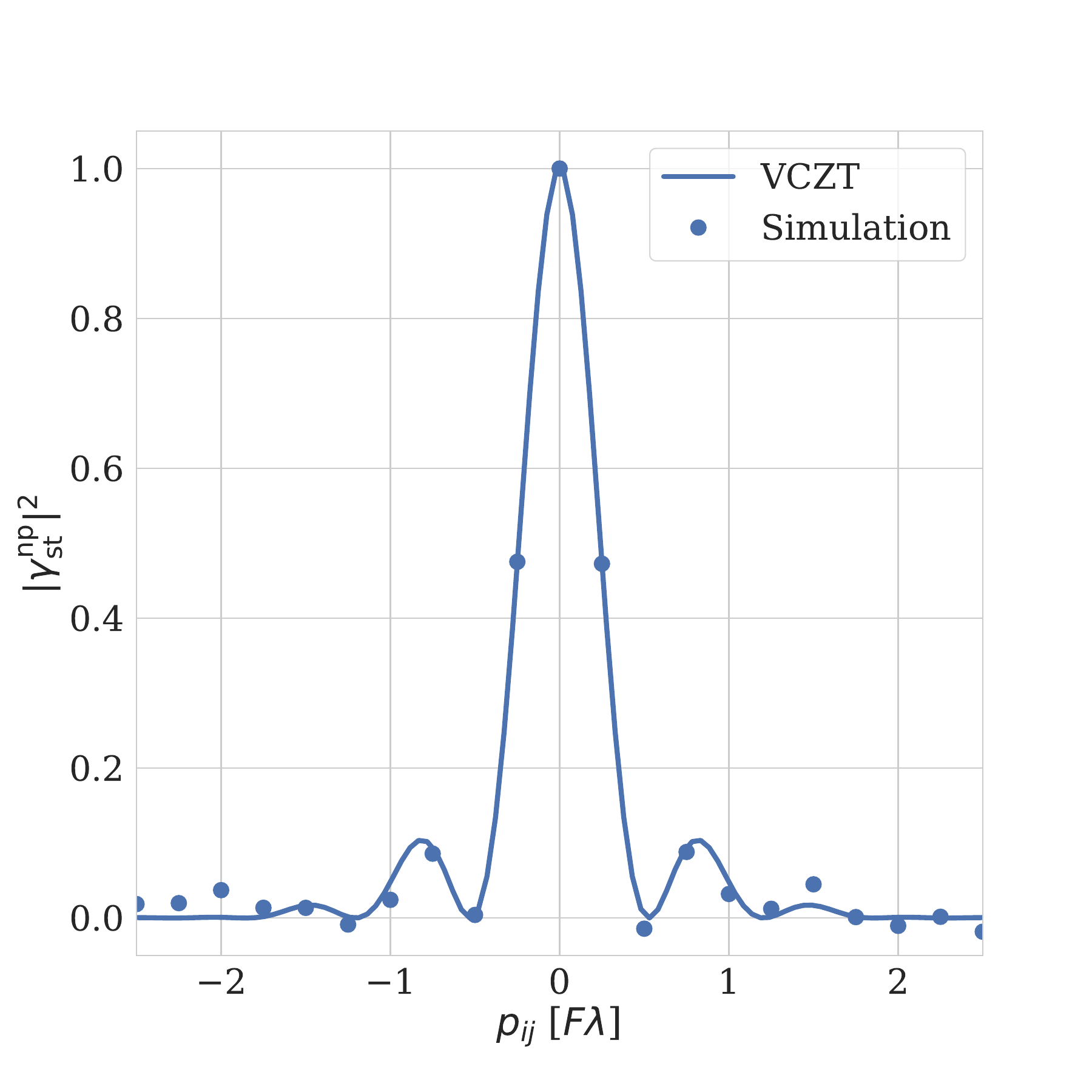}
    \end{subfigure}
    \caption{Simulated correlation patterns at the focal plane due to radiation within the aperture (left) and radiation from the stop (right). The points represent the Monte Carlo simulations described in Sec.~\ref{sec:corr_patterns}, and the lines show the expectation of the van Cittert-Zernicke theorem for an aperture/stop with uniform illumination.}
    \label{fig:hbt_corr_coeff}
\end{figure*}

Figure~\ref{fig:hbt_corr_coeff} shows spatial HBT correlation patterns for both aperture and stop radiation
calculated using their respective VCZT coefficients.
As a demonstration, Fig.~\ref{fig:hbt_corr_coeff} also shows  Monte-Carlo simulations where we directly calculate the (classical) intensity correlation coefficient by propagating fields to detectors $i$ and $j$ from complex Gaussian-random emitters on the stop and from complex Gaussian-random plane-wave modes incident on the aperture.
We set the aperture diameter to $D_{\mathrm{ap}} = 100 \lambda$, the stop's outer diameter to $2 D_{\mathrm{ap}}$, which simulates the case of $\sigma_{\mathrm{ap}} = D_{\mathrm{ap}}/2$, and the f-number to $F = 2.5$. We then simulate 3,000 realizations the aperture and stop fields. 
For radiation from the stop, we integrate Gaussian-random emitters over a grid with cell size $\lambda^2$.\footnote{The chosen $\lambda$ grid spacing corresponds approximately to the coherence length of the simulated thermal radiation and therefore is small enough to represent the stop as a completely incoherent source. See Appendix~\ref{app:partial_coherence} for more details.}
For radiation within the aperture, we simulate 2,500 total plane-wave modes (50 for each of $x$ and $y$ wave-number grids) distributed over the $2\pi\,\mathrm{sr}$ incident solid angle,  which is enough to both resolve the aperture's edges and Nyquist sample the integration
grid.
As expected, the Monte-Carlo simulation yields consistent results with the semi-analytic calculation of HBT coefficients in Eqs.~(\ref{eq:aperture_coherence}) and~(\ref{equ:stop_radiation_coherence}).

Combining the aperture- and stop-radiation components, we obtain the VCZT coefficient for the assumed optical system
\begin{widetext}
\begin{equation}
\begin{split}
    \gamma^{\rm np}_{ij}(\nu)
    & =  \frac{n(T_{\rm (ap)}, \nu)}{n(T_{(i)}, \nu)} \iint\limits_{\sqrt{u^2 + v^2} < D_{\rm ap}/2}
    \!\!\!\!\!\!\!\!\!\!
    \dd u \, \dd v \, |G_i(u, v)|^2 \,
    e^{2\pi i \, p_{ij} u /D_{\rm ap} F \lambda} 
     +
    \frac{n(T_{\rm stop}, \nu)}{n(T_{(i)}, \nu)} \iint\limits_{\sqrt{u^2 + v^2} \geq D_{\rm ap}/2}
    \!\!\!\!\!\!\!\!\!\!
    \dd u \, \dd v \, |G_i(u, v)|^2 \,
    e^{2\pi i \, p_{ij} u /D_{\rm ap} F \lambda}
    \: ,
    \\
    & =  \frac{n(T_{\rm (ap)}, \nu)}{n(T_{(i)}, \nu)}
    \: \eta_{\rm ap} \: \gamma^{\rm np}_{{\rm ap}, ij} (\nu)
     +
    \frac{n(T_{\rm stop}, \nu)}{n(T_{(i)}, \nu)}
    \: (1-\eta_{\rm ap}) \: \gamma^{\rm np}_{{\rm stop}, ij} (\nu)
    \:    ,
\end{split}
\label{equ:vczt_telescope_master_equation}
\end{equation}
\end{widetext}
where $n(T_{(i)}, \nu) \simeq n(T_{(j)}, \nu)$ is defined in Eq.~(\ref{equ:effective_brightness}) and satisfies
\begin{equation}
 n(T_{(i)}, \nu) = 
    n(T_{\rm (ap)}, \nu) \, \eta_{\rm ap} + n(T_{\rm stop}, \nu) \, (1-\eta_{\rm ap}) \; .
\label{equ:effective_brightness_temperature_with_ap_stop}
\end{equation}
Equation~(\ref{equ:vczt_telescope_master_equation})
provides a general prescription to calculate the optical intensity correlation coefficient $|\gamma^{\rm np}_{ij}(\nu)|^2$ for bolometric detectors given a pixel spacing $p_{ij}$ and an optical system parameterized by $G_i(u, v)$, $F$, and $D_{\rm ap}$. In addition, Eqs.~(\ref{equ:flat_illum_approx}) and (\ref{eq:stop_coherence_approx}) can be used as handy approximations to evaluate Eq.~(\ref{equ:vczt_telescope_master_equation}), such as for the sensitivity estimations in Sec.~\ref{sec:sensitivity}.

\subsection{Polarized correlation patterns}
\label{sec:polarized_correlation_patterns}

Following the assumptions for telescope optics in Sec.~\ref{sec:telescope_model} and detector parameters in Sec.~\ref{sec:focal_plane_model}, we show here that the polarized correlation pattern becomes the same as those derived in Sec.~\ref{sec:simple_with_pol}.

We consider detector pixels $i$ and $j$ and assume that each pixel $i$ has two detectors $i1$ and $i2$ with orthogonal polarization angles $\psi_i$ and $\psi_i+\pi/2$ following the beam patterns defined in Eqs.~(\ref{equ:ideal_detector}) and (\ref{equ:ideal_detector2}), respectively.
As described in Sec.~\ref{sec:simple_with_pol}, each infinitesimal thermal source area on the aperture plane emits radiation with two independent Ludwig-3 polarizations $\hat{e}^{L3}_1$
and $\hat{e}^{L3}_2$. This radiation couples to polarized plane-wave modes $\hat{e}^{L3}_1(\theta_i, \phi_i)$ and $\hat{e}^{L3}_2(\theta_i, \phi_i)$ that propagate along direction $(\theta_i, \phi_i)$, and the telescope transforms these plane waves into spherical waves with polarization patterns $\hat{e}_1^{L3}(\theta_{i, \rm pix}, \phi_{i, \rm pix})$ and $\hat{e}_2^{L3}(\theta_{i, \rm pix}, \phi_{i, \rm pix})$, as described in Eq.~(\ref{eq:ludwig3_spherical_optics}).
In the reverse-time sense, the plane-wave modes radiating from detectors $i1$ and $i2$ correspond to plane-wave polarization vectors
\begin{equation}
    \begin{split}
        \hat{e}^{L3}_1(\theta_i, \phi_i) \cos \psi_i & + \hat{e}^{L3}_2(\theta_i, \phi_i) \sin \psi_i \:\:
        \mathrm{and} \\
        -\hat{e}^{L3}_1(\theta_i, \phi_i) \sin \psi_i & + \hat{e}^{L3}_2(\theta_i, \phi_i) \cos \psi_i \: ,
    \end{split}
\end{equation}
respectively, when passing through the aperture. We note here that $\hat{e}_{1 (2)}^{L3}$ does not cross-couple to $\hat{e}^{L3}_{2 (1)}(\theta_i, \phi_i)$, and we assume that our ideal telescope generates no cross polarization.

Given the above assumptions, the polarization degree of freedom introduces a factor of $\cos \psi_i$ and $\sin \psi_i$ for the coupling between detector $i1$ and the $\hat{e}^{L3}_1$ and  $\hat{e}^{L3}_2$ emitters, respectively, with similar behavior for detector $i2$.  
The rest of the calculation proceeds exactly as in Sec.~\ref{sec:simple_with_pol}, following the results in Eqs.~(\ref{equ:mutual_intensity_pol1}), (\ref{equ:vczt_with_pol}), (\ref{equ:qq_covariance}), and (\ref{equ:qq_correlation}) that relate the unpolarized HBT coefficient $\gamma_{ij}^{{\rm np}} (\nu)$ to the Stokes $Q$ coherence as
\begin{equation}
    \gamma_{ij}^{Q, (2)} (\nu)
    = \cos [2(\psi_i - \psi_j)] \, |\gamma_{ij}^{{\rm np}} (\nu)|^2 \:.
    \label{equ:qq_correlation_with_telescope}
\end{equation}

It is worth noting that the results derived here indicate zero correlation between two orthogonal detectors on different pixels $i1$ and $j2$ when $\psi_i = \psi_j$.
This finding contrasts the situation for planar blackbody radiators (or absorbers), where a non-zero correlation structure arises between $\hat{x}$ radiation from one location and $\hat{y}$ radiation from another~\cite{mehta_coherence_1964}.
This difference 
comes from our assumption that the detector beam has an ideal Ludwig-3 polarization pattern with $E/H$-plane symmetry.
For emitters on a surface, modes project onto the $\hat{x}$ the $\hat{y}$ directions with a factor of $\cos \theta_i$ and are therefore not $E/H$-symmetric, hence introducing coherence along the two orthogonal directions.\footnote{One may regard this discrepancy as an artifact of the modal definition, as a fraction of the surface emitter's modes projects to the $\hat{z}$ direction. For the case of ideal Ludwig-3 detectors, all modes are distributed between the two orthogonal detectors, which results in $E/H$-plane symmetry.}

\subsection{VCZT's assumptions and applicability}
The van Cittert-Zernike Theorem (VCZT) relies on a few assumptions, and thus its application has limitations (e.g., see Ref.~\cite{isra.vczt.2017} and references therein).
We now comment on these assumptions and justify the applicability of the VCZT formalism to the intensity coherence calculation for millimeter and submillimeter telescopes.

First, the VCZT in its simplest form assumes scalar waves without a polarization degree of freedom and that the detectors are in the far-field of the source so that the Fraunhofer approximation can be used. That said, our $S$-matrix based formalism accommodates general mode-to-mode coupling and therefore does not rely on these assumptions.
Second, VCZT assumes completely incoherent sources, but our formalism assumes blackbody radiators, which exhibit a coherence length of $\sim \lambda$. Appendix~\ref{app:partial_coherence} shows that this one-wave coherence length can be ignored and therefore that our application of VCZT to blackbody sources is justified.
Third, when considering amplitude coherence, we require $|R_i - R_j| \ll c/\Delta \nu$ to avoid the problem of decoherence. In contrast, decoherence does not occur for intensity coherence as noted in Sec.~\ref{sec:zmu_circuit} and therefore no such bandwidth limit exists for calculations in this paper.
%


\section{Impact of correlations on sensitivity}
\label{sec:sensitivity}

Using Eqs.~(\ref{equ:vczt_telescope_master_equation}), (\ref{eq:correl_1_vs_2}), and (\ref{equ:qq_correlation_with_telescope}), we now investigate the impact of detector-to-detector correlations on instrument sensitivity, which is the primary goal of this paper. As shown in Fig.~\ref{fig:hbt_corr_coeff}, detector outputs can correlate if their pixel pitch is $D_{\mathrm{pix}} < 1.2 F \lambda$, and these correlations will slow noise averaging during coaddition and therefore degrade array sensitivity.\footnote{Strictly speaking, a positive correlation between neighboring pixels can technically improve sensitivity when the angular diameter of interest approaches the pixel spacing projected on the sky (e.g., for Sunyaev Zel'dovich galaxy cluster surveys).}
In this section, we introduce a formalism for mapping speed, which measures the total sensitivity of the detector array, and we inspect the impact of HBT correlations on mapping speed vs.\ pixel size, which is a key metric used for focal plane design.\footnote{The code used to generate plots in this section can be found at \href{https://github.com/chill90/HBT-Correlations}{https://github.com/chill90/HBT-Correlations}.}


\subsection{Mapping speed}
\label{sec:mapping_speed}

We assume that each detector in the imaging array has three noise components: photon shot noise, photon wave noise, and internal noise.
The covariance between detection output ports $i$ and $j$
is
\begin{equation}
    \sigma_{ij}^2 = \sigma_{ij, \rm shot}^2
    + \sigma_{ij, \rm wave}^2
    + \sigma_{ij, \rm int}^2 \: ,
    \label{eq:covariance_matrix}
\end{equation}
and the total variance of the detector array is
\begin{equation}
    \sigma_{\mathrm{arr}}^{2} = \frac{1}{N_{\mathrm{det}}^{2}} \sum_{i, j} \sigma_{ij}^2 \; ,
    \label{eq:total_variance}
\end{equation}
which effectively quantifies the instrument's array-averaged sensitivity.  

To both simplify and clarify the calculations that follow, we assume that all detectors in the array have the same noise properties.
Internal detector noise and photon shot noise cannot correlate between outputs, which allows the covariance to be written as
\begin{equation}
    \sigma_{ij}^2 = \left( \sigma_{\mathrm{shot}}^{2} + \sigma_{\mathrm{int}}^{2} \right) \delta_{ij} + \gamma_{ij}^{(2)}  \sigma_{\mathrm{wave}}^{2} \, ,
    \label{eq:covariance_matrix_uniform_array}
\end{equation}
where $\gamma_{ij}^{(2)}$ is the HBT coefficient and where $\sigma_{\mathrm{shot}}^2$, $\sigma_{\mathrm{wave}}^2$, and $\sigma_{\mathrm{int}}^2$ are the variances of the shot, wave, and internal noise components, respectively, for every detector in the array. While array uniformity is not in general true for real experiments, it is common practice to use the median noise expectation when forecasting instrument performance, making a uniform treatment useful for instrument designers. Additionally, the details of detector-to-detector variation are experiment-dependent and are therefore beyond the scope of this paper.

Given the simplification in Eq.~(\ref{eq:covariance_matrix_uniform_array}) and noting that $\gamma_{ii}^{(2)} = 1$, the total variance of the detector array can be written as
\begin{equation}
    \sigma_{\mathrm{arr}}^{2} = \frac{\sigma_{\mathrm{shot}}^{2} + \sigma_{\mathrm{wave}}^{2} + \sigma_{\mathrm{int}}^{2}}{N_{\rm det}}
    + \frac{\sigma_{\mathrm{wave}}^{2}}{N_{\mathrm{det}}^2} \sum_{i} \sum_{j \neq i} \gamma_{ij}^{(2)}  \: .
    \label{eq:array_noise_terms_separated}
\end{equation}
The first term represents uncorrelated array noise while the second term quantifies noise augmentation due to HBT correlations. Let us further define an array-averaged correlation coefficient across the detector array as
\begin{equation}
    \gamma^{(2)} \equiv \frac{1}{N_{\mathrm{det}}} \sum_{i} \sum_{j \neq i} \gamma_{ij}^{(2)} \, ,
    \label{eq:total_gamma}
\end{equation}
\noindent
which allows us to write the array sensitivity more compactly as
\begin{equation}
    \sigma_{\mathrm{arr}}^{2} = \frac{\sigma_{\mathrm{shot}}^{2} + (1 + \gamma^{(2)}) \sigma_{\mathrm{wave}}^{2} + \sigma_{\mathrm{int}}^{2}}{N_{\mathrm{det}}} \, .
    \label{eq:array_noise_compact}
\end{equation}
\noindent
In this form, the impact of intensity correlations on the detector array noise is reduced to calculating the array-averaged HBT coefficient $\gamma^{(2)}$. In the limit of $\gamma_{ij}^{(2)} \rightarrow 1$, $\gamma^{(2)} \rightarrow (N_{\mathrm{det}} - 1)$ and the array-averaged wave noise is not at all suppressed by detector coaddition. This fact drives the use of interferometers at low frequencies where $n(\nu, T) \gg 1$, where $\sigma_{\mathrm{wave}} \gg \sigma_{\mathrm{shot}}$, and where correlation lengths are long for astronomical sources. In the other limit of $\gamma_{ij}^{(2)} \rightarrow 0$, wave noise averages in the familiar way for uncorrelated measurements $\sigma_{\mathrm{wave}}^{2} / N_{\mathrm{det}}$. It is worth emphasizing that the augmentation of $\sigma_{\mathrm{arr}}^{2}$ by $\gamma^{(2)}$ not only depends on the HBT coefficient $\gamma_{ij}^{(2)}$ but also on the relative contribution of wave noise to that of the other noise terms. As shown in Eq.~(\ref{eq:shot_wave_noise}), $\sigma_{\mathrm{wave}}^{2} \propto n^{2}(\nu, T)$ while $\sigma_{\mathrm{shot}}^{2} \propto n(\nu, T)$, and therefore $\sigma_{\mathrm{wave}}^{2}$ becomes more important at lower frequencies and higher brightness temperatures.

Internal detector noise $\sigma_{\mathrm{int}}$ depends on many factors, including the detector's architecture, thermal noise properties, amplifier noise properties, linearity, and dynamic range, among other things. To remain agnostic to these experiment-specific characteristics, we set $\sigma^{2}_{\mathrm{int}} / \sigma^{2}_{\mathrm{ph}} = 0.1$ hereafter, noting that modern mm-wave observatories aim to be photon-noise dominated.\footnote{For the ubiquitous transition-edge sensor (TES) bolometric detectors~\cite{richards_bolometers_1994,lee_voltage-biased_1997}, internal thermal noise is proportional to the detector's saturation power $P_{\mathrm{sat}} = P_{\mathrm{opt}} + P_{\mathrm{bias}}$, where $P_{\mathrm{opt}}$ is the detected optical power and $P_{\mathrm{bias}}$ is the detector bias power~\cite{mather_bolometer_1982}. Therefore, scaling $\sigma_{\mathrm{int}}$ and $\sigma_{\mathrm{ph}}$ together during experiment design and optimization, while not an exact metric, is well motivated.}

Finally, we quantify the experiment's signal-to-noise using its mapping speed, which is defined as the square ratio of the input signal $S$
to the array-averaged noise $\sigma_{\rm arr}$
\begin{equation}
    MS = \frac{S^{2}}{\sigma_{\mathrm{arr}}^{2}} \propto \frac{N_{\mathrm{det}} \, \eta^{2}}{\sigma_{\mathrm{shot}}^{2} + (1 + \gamma^{(2)}) \, \sigma_{\mathrm{wave}}^{2} + \sigma_{\mathrm{int}}^{2}} \, .
\end{equation}
\noindent
Here, $\eta$ is the optical efficiency of the entire system, which is a product of the detector's quantum efficiency (Eq.~(\ref{equ:det_quantum_efficiency})), the aperture spillover efficiency (Eq.~(\ref{eq:aperture_spill_efficiency})),
and all the other transmission efficiencies, including those of the telescope's optical elements and the atmosphere.
Mapping speed is a powerful measure of an experiment's efficacy, as it is $\propto N_{\mathrm{det}}$ and is therefore analogous to detector yield and observation efficiency.


\subsection{Pixel size optimization}
\label{sec:focal_plane_optimization}

Using the optics and detector assumptions in Sec.~\ref{sec:optical_model} and the VCZT and HBT coefficients in Eqs.~(\ref{equ:vczt_telescope_master_equation}), (\ref{eq:correl_1_vs_2}), and (\ref{equ:qq_correlation_with_telescope}), we now calculate mapping speed vs.\ pixel size. Provided a fixed FOV, or equivalently a fixed focal plane size, decreasing pixel diameter $D_{\mathrm{pix}}$ increases the number of detectors as $N_{\mathrm{det}} \propto D_{\mathrm{pix}}^{-2}$ but decreases aperture spillover efficiency as $\eta_{\mathrm{ap}} \propto \exp [- \left( \pi D_{\mathrm{pix}} / (F \lambda w_{\mathrm{f}}) \right)^{2} / 2]$. These competing effects combine to form a peak in mapping speed vs.\ pixel size, which reveals the optimal packing density.

\begin{figure}[tbp]
    \centering
    \includegraphics[width=0.4\textwidth, trim=2cm 1cm 2.5cm 3.5cm, clip]{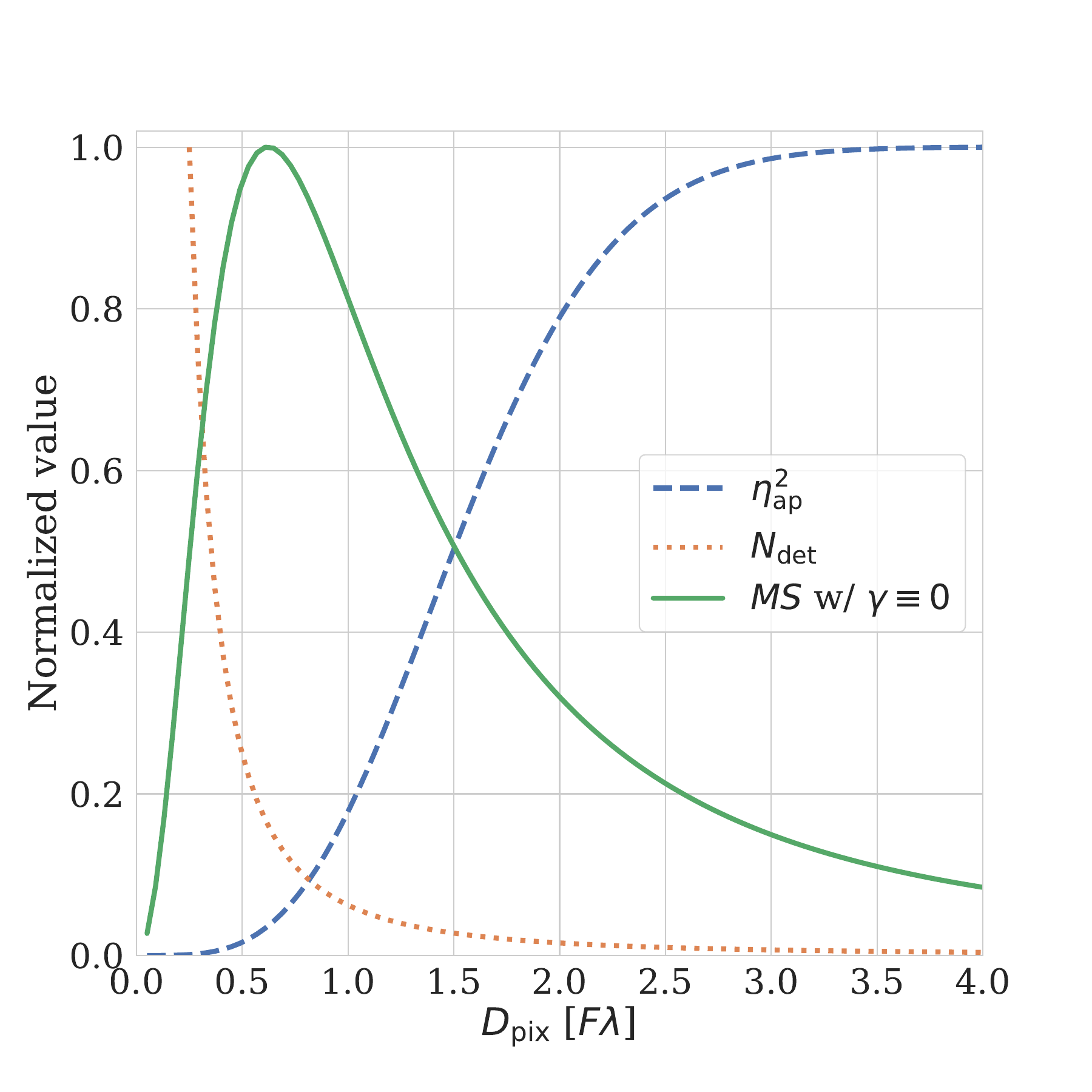}
    \caption{An example of a ``classic'' mapping speed calculation---which ignores the impact of HBT correlations---for a 90~GHz telescope with a 4~K aperture stop. The $MS$ peak arises from the opposing effects of more detectors to average vs. less aperture efficiency with decreasing pixel size. The optimum when ignoring correlations is $\sim 0.7\,F \lambda$, but as shown below, the addition of the HBT coefficient modifies this classic curve.}
    \label{fig:mapping_speed_example}
\end{figure}

An example ``classic'' mapping speed vs.\ pixel size curve~\cite{griffin_relative_2002,2010PhDT.......176A,suzuki_multichroic_2013,Datta2014HornExperiment,Cukierman2018HierarchicalWavelengths}---one for which $\gamma^{(2)} \equiv 0$---of a model 90~GHz instrument with a 4~K stop is shown in Fig.~\ref{fig:mapping_speed_example}. Historically, ground-based CMB experiments have observed at 95, 150, and 220 GHz with $D_{\mathrm{pix}} = 1 \sim 2 F \lambda$~\cite{2010PhDT.......176A,Henning2012Feedhorn-coupledSPTpol,suzuki_multichroic_2013,Thornton2016THEINSTRUMENT,Simon2016TheACTPol,Simon2018TheArray,Posada2018FabricationReceiver,Hui2016BICEP3Performance,Schillaci2020DesignReceiver,Dahal2020TheCharacterization} where the HBT correlation coefficient is small. However, as new readout architectures become available~\cite{dobbs_frequency_2012,dober_microwave_2017} and as CMB experiments push to lower frequencies for improved synchrotron characterization~\cite{Li2018PerformanceArray,Xu2020Two-yearPerformance,Zhang2020CharacterizingArray}, focal planes with $D_{\mathrm{pix}} \lesssim 1.2 F \lambda$ are becoming increasingly practical. Therefore, the impact of HBT correlations on mapping speed is of interest to upcoming mm-wave experiments, such as CMB-S4~\cite{abitbol_cmb-s4_2017}. In this section, we calculate HBT-modified mapping speed vs.\ pixel size curves for our model telescope at several observation frequencies, and we discuss the results.

We assume that our telescope is ground-based with cryogenically cooled optics, infrared filters, and sub-Kelvin detectors. As shown in Eq.~(\ref{equ:t_aperture_definition}), radiation at the aperture plane can be summed over all sky-side sources and represented by a single effective brightness temperature $T_{(\mathrm{ap})}$. To simplify and generalize the following analysis, we assume only three sources viewed through the aperture: the CMB with $T_{\mathrm{CMB}}$, the atmosphere with $T_{\mathrm{atm}}$, and telescope optics with $T_{\mathrm{tel}}$.
The telescope's effective temperature $T_{\mathrm{tel}}$ can vary considerably depending on the specifics of the mirrors, ground shield, cryostat window, and anti-reflection coatings, but as an example, we assert a default configuration where $T_{\mathrm{tel}} = 10$~K. We also assert the stop's physical temperature to be
$T_{\mathrm{stop}} = 4$~K and that each detector's quantum efficiency is $\eta_{\mathrm{det}} = 0.8$. 
To model power due to atmospheric emission, we assume that the telescope observes from the the Chajnantor Plateau in the Atacama Desert of Chile, and we use the AM model~\cite{2019zndo...3406483P} to generate the atmosphere's effective brightness temperature and transmittance at 1~mm precipitable water vapor (PWV) and 50~deg elevation above the horizon.

\begin{figure*}[tbp]
    \begin{subfigure}
        \centering
        \includegraphics[width=0.45\textwidth, trim=0.1cm 1.3cm 2.1cm 3cm, clip]{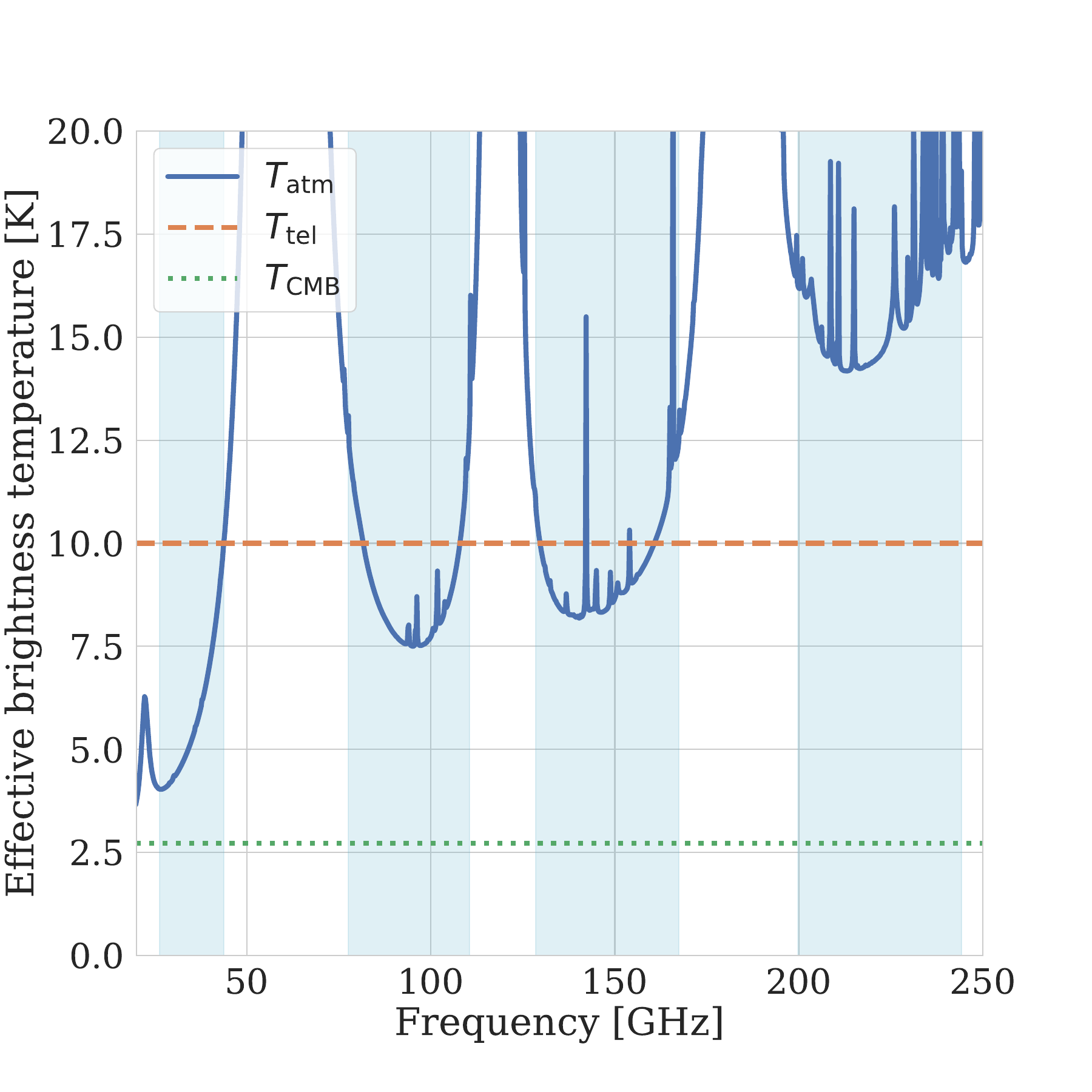}
    \end{subfigure}
    \hspace{0.05\textwidth}
    \begin{subfigure}
        \centering
        \includegraphics[width=0.43\textwidth, trim=0.5cm 1.3cm 2.4cm 2.5cm, clip]{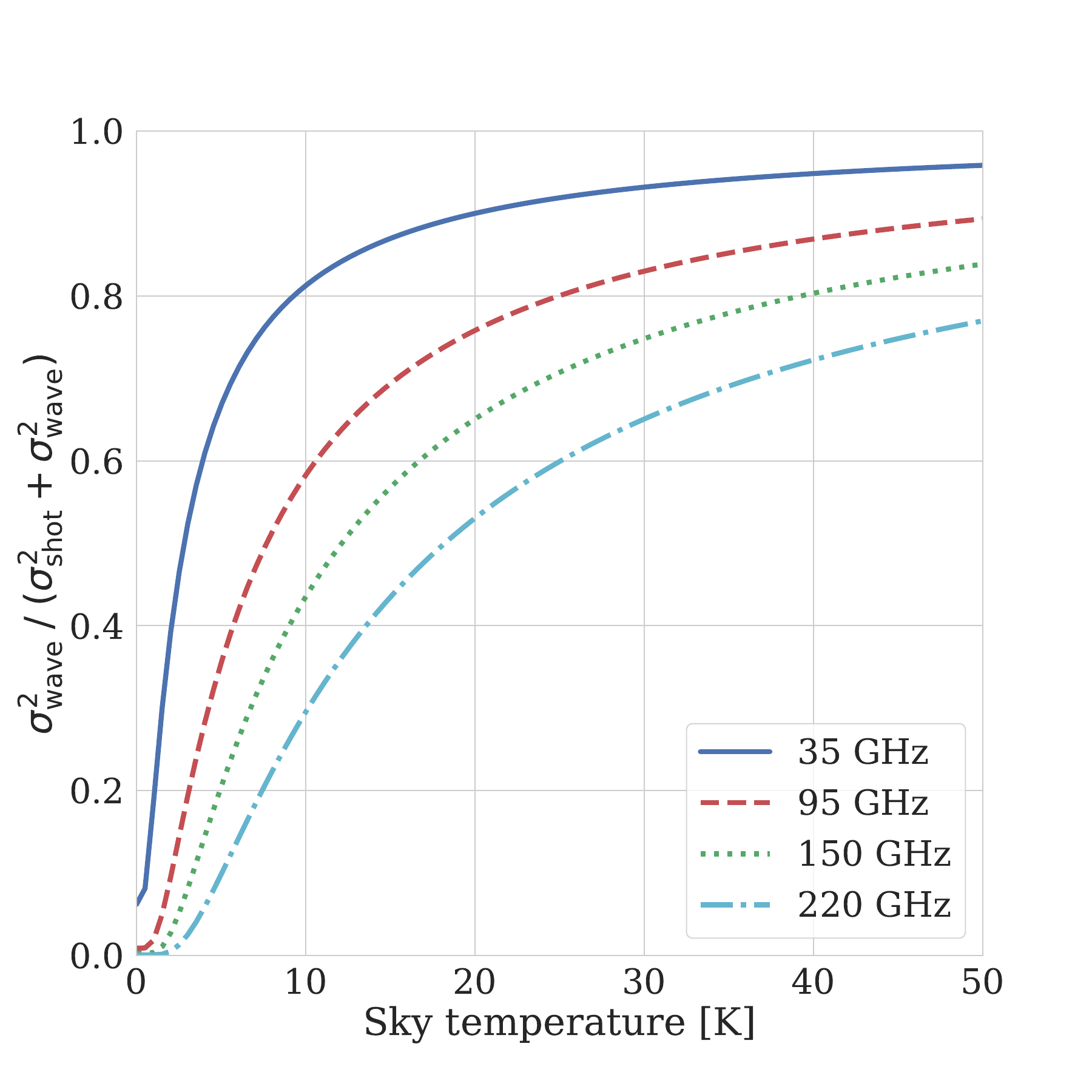}
    \end{subfigure}
    \caption{The assumed observation bands plotted over the assumed CMB, atmosphere, and telescope temperatures (left), and the wave-noise fraction vs.\ sky temperature for each band at $D_{\mathrm{pix}} = 3 F \lambda$, where $\eta_{\mathrm{ap}} > 0.95$ (right). 
    }
    \label{fig:observation_bands}
\end{figure*}

Given this telescope + sky model, we consider four top-hat observation bands centered at (35, 95, 150, 220)~GHz with bandwidths of (17, 33, 39, 44)~GHz. The chosen bands, along with $T_{\mathrm{CMB}}$, $T_{\mathrm{atm}}$, and $T_{\mathrm{tel}}$, are shown in Fig.~\ref{fig:observation_bands}. The channels are chosen to fit within atmospheric windows and are similar to those of existing Atacama instruments. Additionally, Fig.~\ref{fig:observation_bands} shows the bunching fraction for each frequency band vs.\ sky temperature at $D_{\mathrm{pix}} = 3 F \lambda$, where $\eta_{\mathrm{ap}} > 0.95$. As discussed in Sec.~\ref{sec:mapping_speed}, while lower-frequency modes tend to have larger occupation numbers, the sky is brighter at higher frequencies, and therefore correlations substantially impact sensitivity in all four bands.

\begin{figure*}[tbp]
    \centering
    \includegraphics[width=1.0\linewidth, trim=0cm 0cm 0cm 0cm, clip]{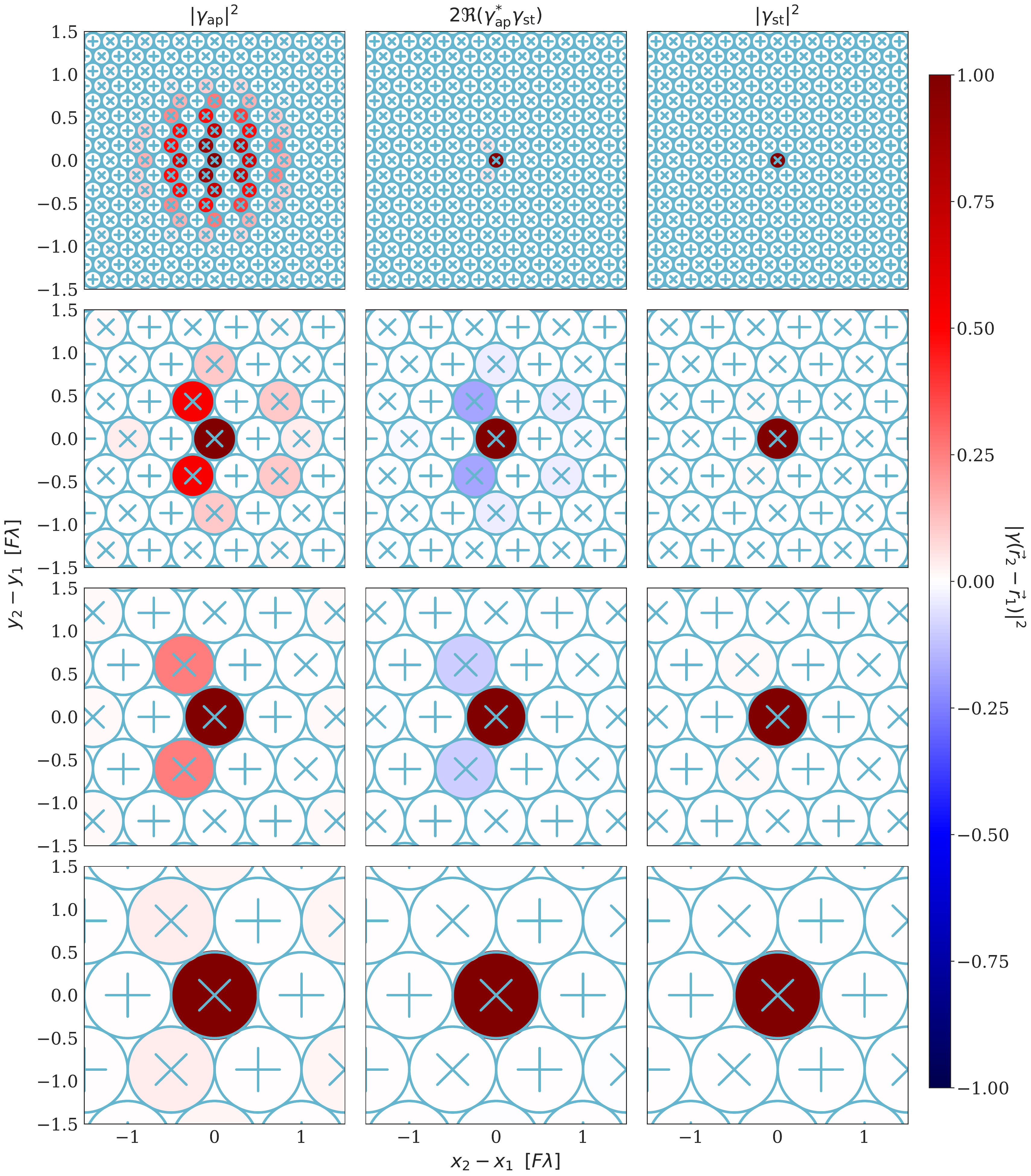}
    \caption{The Stokes Q HBT coefficient $\gamma^{Q, (2)}_{ij} = \cos 2 (\psi_i-\psi_j) \, \gamma^{(2)}_{ij}$ given pixel pitches $D_{\mathrm{pix}} =$ (0.2, 0.5, 0.7, 1.0) $F \lambda$ from top to bottom for radiation from within the aperture (left), from the stop (right), and their cross term (middle). When the pixel pitch is $\lesssim F \lambda$, the detectors oversample the spatial modes on the focal plane, giving rise to intensity correlations between nearby pixels.}
    \label{fig:pixel_packing_contours}
\end{figure*}

Fig.~\ref{fig:pixel_packing_contours} shows an example of four pixel pitch scenarios $D_{\mathrm{pix}} = (0.2, 0.5, 0.7, 1.0) \, F \lambda$ given the hex packing described in Sec.~\ref{sec:focal_plane_model}, and the contour shows the Stokes $Q$ HBT coefficient $\gamma^{Q, (2)}_{ij}$ due to both aperture and stop radiation. As expected, the degree of correlation is a two-dimensional version of the VCZT curves in Fig.~\ref{fig:hbt_corr_coeff}. The effect of pushing $D_{\mathrm{pix}} \lesssim F \lambda$ is for the detected modes (the pixel apertures) to oversample the input modes (the $\gamma^{Q, (2)}_{ij}$ contour), giving rise to correlated noise between nearby detectors.

\begin{figure*}[tbh]
    \centering
    \includegraphics[width=0.8\linewidth, trim=0cm 0cm 0cm 0cm, clip]{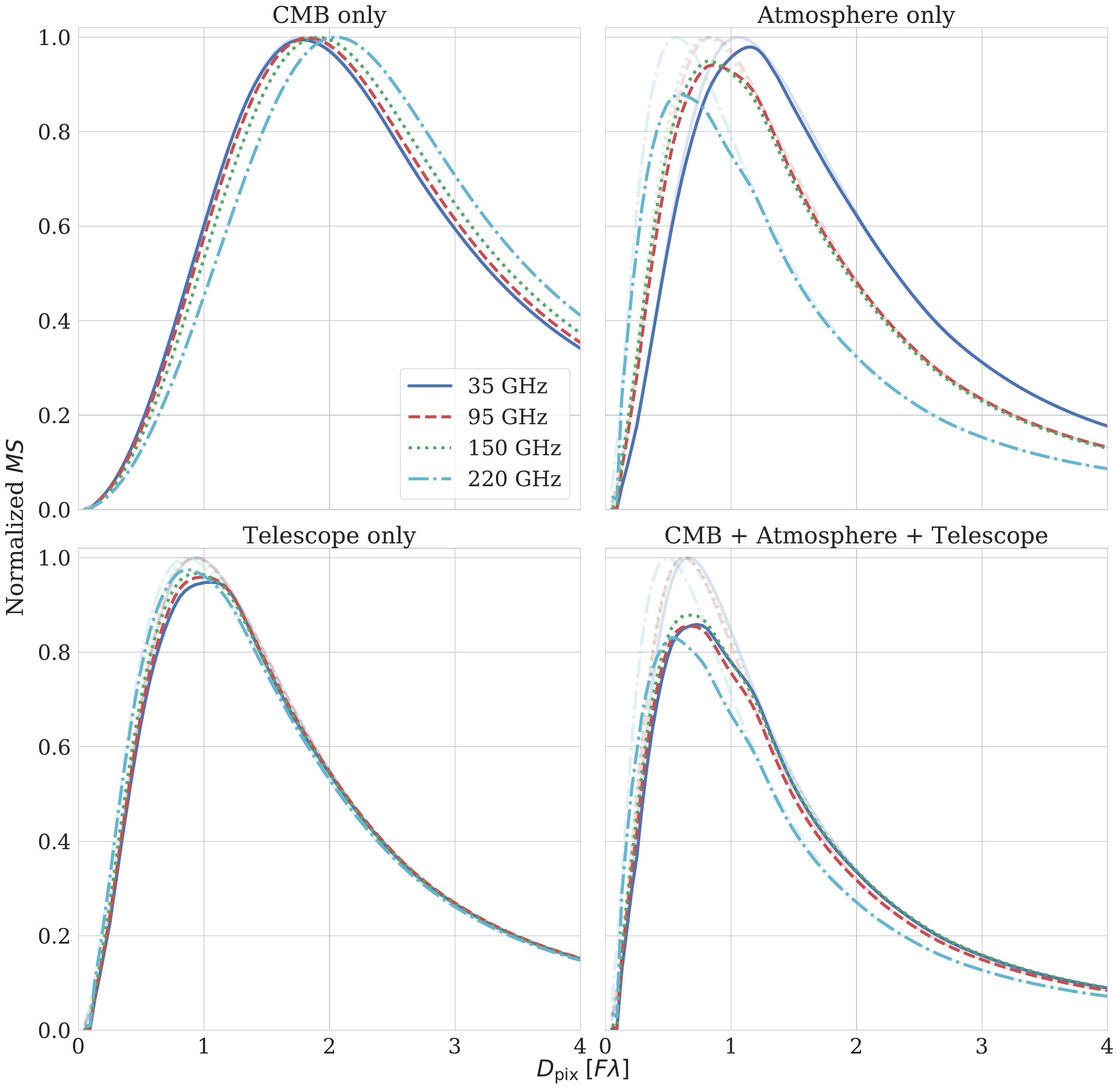}
    \caption{Stokes $Q$ mapping speed vs.\ pixel size in each frequency band for four sets of external sources illuminating the aperture: the CMB only with $T_{\mathrm{CMB}} = 2.725$ K (top left), the atmosphere only assuming 1 mm PWV and 50 deg elevation at the Chajnantor Plateau (top right), the telescope emission only assuming $T_{\mathrm{tel}} = 10$ K (bottom left), and all three sources combined (bottom right). We assume that each detector's quantum efficiency is $\eta_{\mathrm{det}} = 0.8$ and that $\sigma_{\mathrm{int}}^{2} = 0.1 (\sigma_{\mathrm{shot}}^{2} + \sigma_{\mathrm{wave}}^{2})$. Each band is normalized to the peak of its classic curve, which is shown as faded lines. Therefore, the opaque curves represent the achievable mapping speed when $\gamma^{(2)} \neq 0$ with respect to the maximum of the $\gamma^{(2)} \equiv 0$ case. Pixel size is plotted in units of $F \lambda$, where $\lambda$ is the mean wavelength in each band.}
    \label{fig:mapping_speed}
\end{figure*}

When calculating Stokes $Q$ mapping speed, we sum the HBT coefficients in Fig.~\ref{fig:pixel_packing_contours} over a 4~$F \lambda$ radius to find
\begin{equation}
    \gamma^{Q, (2)}_{i} \equiv \sum_{j \neq i} \gamma_{ij}^{Q, (2)} \:,
\end{equation}
noting that because each pixel has two orthogonal polarimeters, $\left| \gamma_{ij} \right|^{2}$ vanishes for half of all $(i, j)$ output pairs.
The impact of HBT correlations becomes roughly twice when considering mapping speed for measurements of intensity or Stokes $I$.
We then assume that this sum applies to all detectors on the focal plane such that $\gamma^{Q, (2)} \simeq \gamma^{Q, (2)}_{i}$. This treatment ignores the fact that edge pixels have fewer neighbors than internal ones, which is a reasonable approximation for focal planes of large area. For small or moderately-sized detector arrays, the fraction of edge pixels may become important, but such details are experiment-dependent and are therefore beyond the scope of this paper. Fig.~\ref{fig:mapping_speed} shows HBT-impacted mapping speed vs.\ pixel size curves for each observation band, normalized to the peak of the ``classic'' curve for which $\gamma_{ij}^{Q, (2)} \equiv 0$. Three additional mapping speed curves with in-aperture loading from only the CMB, only the atmosphere, and only the telescope are also plotted to demonstrate the dependence of HBT correlations on various source temperatures.

There are several features in Fig.~\ref{fig:mapping_speed} that are worth noting explicitly. Firstly, the impact of HBT correlations depends on source temperature and is most pronounced in the presence of a brightly illuminated aperture. This effect is most clearly seen when contrasting the CMB and atmosphere, especially at 220~GHz where the CMB's photon occupation number is falling while that of the atmosphere is rising. Secondly, while HBT correlations impact curve shape most prominently at low frequencies, the atmosphere is brighter at higher frequencies, inducing a similar HBT suppression across all bands. Thirdly, the mapping speed peak is located at a slightly larger $D_{\mathrm{pix}}$ than that of the classic curves at 35 and 95~GHz but resides at a similar location to that of the classic curves at 150 and 220~GHz. This effect arises because the stop is significantly fainter than the sky at 150 and 220~GHz, and therefore as $D_{\mathrm{pix}}$ falls below $1.2 F \lambda$, $(1 + \gamma^{(2)})\sigma_{\mathrm{wave}}^2$ decreases less rapidly than $N_{\mathrm{det}} \eta_{\mathrm{ap}}^{2}$. Lastly, the impact of correlations on mapping speed starts to become most important when $D_{\mathrm{pix}} < 1.2 F \lambda$, but there also percent-level impacts at larger spacings, which correspond to local maxima in the aperture/stop VCZT patterns, as shown in Fig.~\ref{fig:hbt_corr_coeff}.
This effect gets smoothed out when averaging the HBT coefficients across each channel's finite bandwidth. Regardless of the input assumptions in this section, the mapping-speed gain by undersized pixels when $\gamma^{(2)} \neq 0$ is suppressed compared to the classic $\gamma^{(2)} \equiv 0$ case, especially for ground-based telescopes.


\section{Implications for experiment design}
\label{sec:implications_for_experiment_design}

\begin{figure*}[tbp]
    \centering
    \includegraphics[width=0.8\textwidth, trim=0cm 0cm 0cm 0cm, clip]{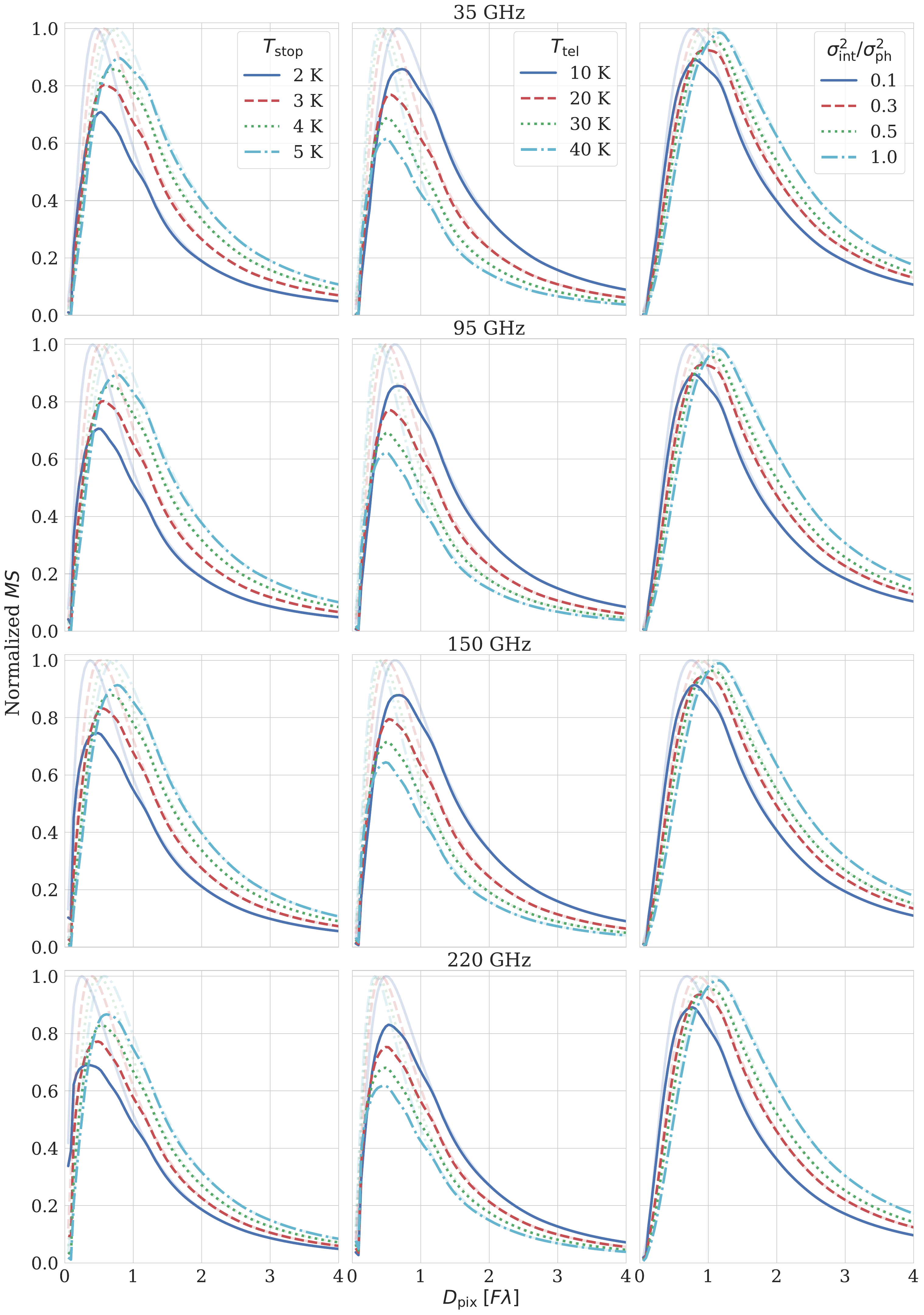}
    \caption{The impact of stop temperature $T_{\mathrm{stop}}$ (left column), telescope temperature $T_{\mathrm{tel}}$ (middle column), and detector internal noise $\sigma_{\mathrm{int}}^{2}$ (right column) on Stokes $Q$ mapping speed vs.\ pixel size in the presence of HBT correlations for each observation band (rows). These parameters are among many that vary between experiments, and we include them here as a reference for focal plane designers. The default parameters from Fig.~\ref{fig:mapping_speed} of $T_{\mathrm{stop}} = 4$ K, $T_{\mathrm{tel}} = 10$ K, and $\sigma_{\mathrm{int}}^{2} / \sigma_{\mathrm{ph}}^{2} = 0.1$ are assumed when not being swept.}
    \label{fig:mapping_speed_dependencies}
\end{figure*}

As shown in Fig.~\ref{fig:mapping_speed}, HBT correlations both modify the optimal pixel packing density and suppress the achievable mapping speed with respect to the ``classic'' $\gamma^{(2)} \equiv 0$ calculation. However, the degree of modification depends on a plethora instrument details, including internal detector noise, observation site and conditions, stop temperature, detector efficiency, telescope optical throughput, and extraneous noise sources, such as electromagnetic interference, vibrational pickup, and detector nonidealities. A more comprehensive handling of correlations within a more general experiment is available via the BoloCalc sensitivity calculator~\cite{hill_bolocalc_2018}, but in this section, we sweep a few parameters in our model telescope to serve as a quick reference for focal plane designers. The results of these calculations are shown in Fig.~\ref{fig:mapping_speed_dependencies}.

The first column of Fig.~\ref{fig:mapping_speed_dependencies} shows mapping speed vs.\ pixel size for various stop temperatures $T_{\mathrm{stop}} = (2, 3, 4, 5)$~K. As stop temperature decreases, so does photon loading due to stop spillover, which in turn favors smaller pixels. A colder stop also suppresses the relative contribution of $\sigma_{\mathrm{wave}}$, especially at higher frequencies, modulating the slope of the mapping speed curve below $D_{\mathrm{pix}} = 1.2 F \lambda$. The second column of Fig.~\ref{fig:mapping_speed_dependencies} shows mapping speed vs.\ pixel size for various telescope temperatures $T_{\mathrm{tel}} = (10, 20, 30, 40)$~K. As telescope temperature increases, so too does the photon load within the aperture, which in turn favors smaller pixels. In addition, brighter aperture radiation increases $\sigma_{\mathrm{wave}}$ and hence also increases the HBT suppression. The third column of Fig.~\ref{fig:mapping_speed_dependencies} shows mapping speed vs.\ pixel size in the presence of a constant internal detector noise $\sigma_{\mathrm{int}}^{2} = (0.1, 0.5, 0.7, 1.0) \times \sigma_{\mathrm{ph}}^{2}$ at $D_{\mathrm{pix}} = 1.2 F \lambda$, where $\sigma_{\mathrm{ph}}^{2} \equiv (\sigma_{\mathrm{shot}}^{2} + \sigma_{\mathrm{wave}}^{2})$. 
As $\sigma_{\mathrm{int}}$ increases with respect to $\sigma_{\mathrm{ph}}$, larger pixel sizes are favored to improve signal strength via an increased $\eta_{\mathrm{ap}}$. Simultaneously, the impact of HBT correlations is reduced due to a smaller relative contribution of $\sigma_{\mathrm{wave}}$ and due to an optimum $D_{\mathrm{pix}} \gtrsim 1.2 F \lambda$ where $\gamma^{(2)}$ is small.

These few examples only graze the rich topic of focal plane optimization, and we leave a more comprehensive discussion of experiment-specific applications to other publications. Nonetheless, regardless of the context, HBT correlations should be considered when designing dense focal planes, especially for ground experiments where the sky and telescope brightness temperatures are substantially larger than that of the CMB.


\section{Conclusion}
\label{sec:conclusion}

We have presented a theoretical formalism for photon noise correlations by extending the quantum optical circuit-based model in Zmuidzinas~\cite{zmuidzinas_thermal_2003} to a free-space classical model using the optical equivalence theorem of Glauber and Sudarshan~\cite{sudarshan_equivalence_1963}. We have used this formalism to estimate the Hanbury Brown-Twiss (HBT) coefficient~\cite{brown_apparent_1952,brown_lxxiv_1954,brown_correlation_1956} in a simplified telescope optical system, and we have shown that these simulations match the expectation of the van Cittert-Zernike theorem (VCZT)~\cite{vanCittert1934DieEbene,Zernike1938TheProblems}. This equivalence allows the HBT coefficient to be calculated with only a knowledge of the radiation intensity profile at the aperture plane.

We then uniformly illuminated our model telescope with blackbody sources representative of radiation from the CMB, atmosphere, and telescope, and we have calculated the impact of HBT correlations on experiment mapping speed vs.\ pixel size within observation bands centered at 35, 95, 150, and 220~GHz. Acknowledging that sensitivity calculations have many inputs and assumptions, we have further discussed three useful variations to the simplified instrument---stop temperature, telescope temperature, and internal detector noise---and showed how each parameter modulates the HBT-modified mapping speed curves. This work builds on an initial discussion by Padin~\cite{padin_mapping_2010} and formalizes the calculation of photon noise correlations between detector pixels in millimeter and sub-millimeter telescopes for astronomy. The presented formalism and results are useful to GHz focal plane designers, especially as emerging readout technologies enable the deployment of dense detector arrays.


\section*{Acknowledgement}
Work at LBNL is supported by the U.S. Department of Energy, Office of Science, Office of High Energy Physics under contract No. DE-AC0205CH11231.  We acknowledge the support by JSPS Grant Number JP19K21873.  We thank our Simons Array and Simons Observatory colleagues for fruitful discussions on CMB telescope designs, and we thank Masahito Ueda for teaching us some of the basics of quantum statistical mechanics.


\appendix


\input{app_thermal_photon_density_matrix}

\input{app_partial_coherence}

\input{app_flat_illumination_vs_general}



\bibliography{charlie_mendeley_rev2,additional_bibs}


\end{document}

%% file: app_thermal_photon_density_matrix.tex
\section{Thermal photon density matrix}
\label{app:thermal_density_matrix}

The statistical state, or the mixed state, of photons can be described using density matrix $\hat{\rho}$. We consider a single-mode photon state, where the state is single-moded in both spatial and frequency domains as well as in polarization state.  Using creation and annihilation operators $a^\dagger$ and $a$, respectively, the density matrix can be written as that of a Bose-Einstein distribution
\begin{equation}
  \hat{\rho} = \frac{e^{-\gamma a^\dagger a}}{\mathrm{Tr}\left(e^{-\gamma a^\dagger a}\right)}
  = \sum_n p(n; \bar{n}) \left|n\right> \left< n\right| \:,
\end{equation}
with
\begin{equation}
  \gamma \equiv \frac{h \nu}{k_\mathrm{B} T} \quad \mathrm{and} \quad 
  p(n; \bar{n}) \equiv \frac{1}{1+\bar{n}}\left(\frac{\bar{n}}{1+\bar{n}}\right)^n \:.
\end{equation}
Here, $\{ \left|n\right> \}$ is the Fock state and $\bar{n}$ is a mean occupation number.  We consider a detection process whose integration time $\tau$ (the inverse of sampling rate) is significantly longer than the coherence time
$\tau_c \equiv 1/\Delta \nu$, where $\Delta \nu$ is the detection bandwidth.
This is usually the case for a CMB instrument, where $\Delta \nu \sim \mathcal{O}(10\,\mathrm{GHz})$ and $\tau \sim \mathcal{O}(10\,\mathrm{msec})$.
In this situation where $\tau \gg \tau_c$, the mean occupation number $\bar{n}$ can be written as
\begin{equation}
  \bar{n} = n(T, \nu) = \frac{1}{e^\gamma - 1} \:.
\end{equation}

We can now rewrite the density matrix in terms of Glauber's coherent state $\left|\alpha \right>$.  The coherent state is written as
\begin{equation}
\left| \alpha \right>
 \equiv e^{\alpha a^\dagger - \alpha^* a} \left| 0 \right>
\end{equation}
and satisfies
\begin{equation}
a \left| \alpha \right> = \alpha \left| \alpha \right>
  \:.
\end{equation}
The density matrix can be rewritten in the  Glauber-Sudarshan $P$ representation~\cite{sudarshan_equivalence_1963,glauber_coherent_1963} as
\begin{equation}
 \hat{\rho} = \int \! d^2\alpha \: p_g(\alpha; \bar{n}) 
   \left| \alpha \right> \left<\alpha \right|
   \label{equ:rho_in_coherent}
\end{equation}
with
\begin{equation}
   p_g(\alpha; \bar{n}) \equiv \frac{1}{\pi \bar{n}} \exp \left(-\frac{|\alpha|^2}{\bar{n}}\right) \:,
\end{equation}
where the integral is over the entire complex plane.
Here, the complex amplitude $\alpha$ follows a Gaussian distribution $p_g(\alpha; \bar{n})$, in agreement with the expectation in the classical limit.
The photon counting of a coherent state follows a Poisson distribution as
\begin{equation}
  \Bigl| \bigl< n \bigl| \alpha \bigr> \Bigr|^2
    = \exp \left(-|\alpha|^2\right) \frac{|\alpha|^{2n}}{n!} \:.
   \label{equ:coherent_state_poisson}
\end{equation}
Eqs.~(\ref{equ:rho_in_coherent}) and (\ref{equ:coherent_state_poisson}) immediately lead to a special case of Mandel's formula~\cite{Mandel_1958,Mandel_1959}
\begin{equation}
\left< n \right| \hat{\rho} \left| n \right>
  = \int_0^\infty \! dW \: e^{-W} \frac{W^{n}}{n!} \:\: \frac{e^{-W/\bar{n}} }{\bar{n}}
  = p(n; \bar{n}) \:.
\end{equation}

%% file: app_partial_coherence.tex
\section{Partial Coherence of Sources}
\label{app:partial_coherence}
In this paper, we assume complete incoherence between two source elements that are physically apart from one another.  However, it is known that blackbody sources have finite correlation at the length scale of a wavelength~\cite{mehta_coherence_1964,carter_coherence_1975,baltes_spectral_1976,steinle_radiant_1977}.
The effect of source coherence on the applicability of VCZT for
quasihomogeneous sources, whose spatial intensity variations are slow compared to its coherence length, are extensively discussed in literature~\cite{wolf_angular_1975,wolf_radiometric_1976,carter_coherence_1977,friberg_radiation_1982}.
Thus, it is worthwhile to clarify the assumptions in this paper regarding source coherence. In the end, we find that the assumption of completely incoherent sources is a good approximation for telescope systems relevant to our discussion.
We first consider source coherence for the simple case presented in Sec.~\ref{sec:simple_example_wo_pol}, which is readily comparable to examples in the literature.  We then discuss source coherence in the general formalism of Sec.~\ref{sec:zmu_circuit}.

Equation~(\ref{equ:simple_mutual_intensity_vczt}) in combination with (\ref{equ:gamma1_in_circuit_model}) constitutes VCZT
of a completely incoherent source
for the simple geometry in Fig.~\ref{fig:simplest_HBT}.  For sources with partial coherence (e.g., see Ref.~\cite{isra.vczt.2017}), the first-order coherence is
\begin{equation}
    \Gamma^{(1)}_{ij}
    \propto
    \iint_\sigma \dd^2 \vec{r}
    \iint_\sigma \dd^2 \vec{r}' \, 
    \gamma_c (|\vec{r} - \vec{r}'|) \,
    e^{2\pi i \bar{ \nu} (R'_j - R_i) / c}
    \: ,
    \label{equ:general_coherence_vczt}
\end{equation}
with
\begin{equation}
R_i \equiv |\vec{r}_i-\vec{r}| \quad \, \quad \quad
 R_j' = |\vec{r}_j-\vec{r}'| \:,
\end{equation}
where $\vec{r}_i$ and $\vec{r}_j$ are the positions of detectors $i$ and $j$, respectively, and where $\gamma_c (|\vec{r} - \vec{r}'|)$
is the coherence of the field between locations
$\vec{r}$ and $\vec{r}'$
on the source surface $\sigma$.
Here, we assume that the source's coherence is statistically isotropic and thus that the source coherence function can be written as $\gamma_c (\vec{r}, \vec{r}') = \gamma_c (|\vec{r} - \vec{r}'|)$.
Complete incoherence of the source corresponds to the limit of 
$\gamma_c (|\vec{r} - \vec{r}'|) \rightarrow \delta (\vec{r} - \vec{r}')$
and we immediately find that Eq.~(\ref{equ:general_coherence_vczt})
reduces to Eq.~(\ref{equ:simple_mutual_intensity_vczt}) in such a limit.

We define a coherence length $R_c$ such that $\gamma_c (|\vec{r} - \vec{r}'|) \simeq 0$ for $|\vec{r} - \vec{r}'| > R_c$.  For blackbody radiators, $R_c \sim \lambda$ since we assume detectors with a limited detection band.\footnote{See, e.g., Ref.~\cite{mehta_coherence_1964} for consideration without a band limit.}  Focusing on cases where $R_c$ is significantly smaller than the source size, Eq.~(\ref{equ:general_coherence_vczt}) can be approximated as
\begin{align}
    \Gamma^{(1)}_{ij} & \propto
    \iint_\sigma \dd^2 \vec{r}
    \iint_{|\vec{\Delta}| \leq R_c} \! \! \! \! \! \!  \! \! \! \! \! \!
        \dd^2 \vec{\Delta} \,\, 
    \gamma_c (|\vec{\Delta}|) \,
    e^{2\pi i \bar{ \nu} (R'_j - R_i) / c}
    \\
    &     \simeq 
    \iint_\sigma \dd^2 \vec{r} \,
    e^{-2\pi i \bar{ \nu} (R_j - R_i) / c}
    \, \, \mathcal{J}(\theta_r)
    \label{equ:arbi_coherence_length_approx}
\end{align}
with
\begin{equation}
    \begin{split}
    \mathcal{J}(\theta_r)
    & \equiv 
    \iint_{|\vec{\Delta}| \leq R_c} \! \! \! \! \! \!  \! \! \! \! \! \!
        \dd^2 \vec{\Delta} \,\, 
    \gamma_c (|\vec{\Delta}|) \,
    e^{2\pi i \frac{\bar{ \nu}}{c} \frac{\vec{r} - \vec{r}_j}{R_j} \cdot \vec{\Delta}}
    \\
    & = \int_0^{2\pi} \!\! \!\! \dd \phi \int_{0}^{R_c}  \! \! \! \! \! \!
            s \, \dd s  \,\, 
    \gamma_c (s) \,
    e^{-2\pi i \sin \theta_r \cdot \cos \phi \cdot s/ \bar{\lambda}}
    \: ,
    \end{split}
\end{equation}
where $\theta_r$ is the angle between $\vec{r}-\vec{r}_j$ and the source plane's normal vector.
The expression in Eq.~(\ref{equ:arbi_coherence_length_approx}) becomes equivalent to 
Eq.~(\ref{equ:simple_mutual_intensity_vczt})
when $\mathcal{J}(\theta_r)$
can be regarded as a constant function of $\vec{r}$.

To evaluate $\mathcal{J}(\theta_r)$, we consider a blackbody surface source that may not be centered at $x=y=0$, and we assume that the solid angle of the source is small compared to $\pi$.  We consider two regions of the parameter space depending on the source's position.  Region 1 is when $\theta_r \ll 1$.  In this case,
\begin{equation}
\left| \sin \theta_r \frac{|\vec{\Delta}|}{\lambda}\right|
 \leq |\sin \theta_r| \frac{R_c}{\lambda} \ll 1 \:,
\end{equation}
and thus $\mathcal{J}(\theta_r) \simeq \mathcal{J}(0)$ and Eq.~(\ref{equ:arbi_coherence_length_approx})
becomes equivalent to Eq.~(\ref{equ:simple_mutual_intensity_vczt}).
Region 2 is when $\theta_r \gtrsim 1$.  In this case, $\theta_r$ is approximately constant across the source $\sigma$, and thus 
$\mathcal{J}(\theta_r) \simeq \mathcal{J}(\bar{\theta}_r)$
where $\bar{\theta}_r$ is the mean of $\theta_r$ for the source $\sigma$. The factor $\mathcal{J}(\bar{\theta}_r) / \mathcal{J}(0)$ maps to the radiance reduction due to source coherence, which shows up in $\Gamma^{(1)}_{ii}$ and $\Gamma^{(1)}_{jj}$ as well.  Thus, the normalized amplitude coherence $\gamma_{ij}$ remains identical to the case of a completely incoherent source, which is consistent with results presented in the literature~\cite{carter_coherence_1977}. When the source comprises Lambertian blackbody emitters,
the radiance satisfies $\mathcal{J}(\bar{\theta}_r) / \mathcal{J}(0) \simeq 1$ by construction.

In summary, for a source with the geometry assumed in Sec.~\ref{sec:simple_example_wo_pol}, partial coherence between spatially independent blackbody radiators
can be neglected and Eq.~(\ref{equ:simple_mutual_intensity_vczt}) is a good approximation.

We now look to Sec.~\ref{sec:zmu_circuit}, which considers a more general, VCZT-free formalism.  For quasihomogeneous sources with partial coherence, in contrast to completely incoherent sources, the Kronecker delta $\delta_{km}$ in Eq.~(\ref{equ:ak_am_correlation}) is replaced by a source coherence function $\gamma_{c, km}$.  The mutual intensity is then expressed as
\begin{equation}
    B_{ij}(\nu) = \sum_\sigma \sum_{k, m \in \sigma} S^{*}_{ik} (\nu) S_{jm} (\nu) \, \gamma_{c, km}^\sigma \, n(T_\sigma,\nu) \:, \label{equ:mutual_intensity_partial_coherence}
\end{equation}
where the index $\sigma$ denotes each quasihomogeneous source with a temperature $T_\sigma$, and $\gamma_{c, km}^\sigma$ is the coherence function of the source.
Equation~(\ref{equ:mutual_intensity_partial_coherence}) is the generalized version of Eq.~(\ref{equ:general_coherence_vczt}).
Similarly to Eq.~(\ref{equ:arbi_coherence_length_approx}), we can decompose Eq.~(\ref{equ:mutual_intensity_partial_coherence}) as
\begin{equation}
    B_{ij}(\nu) \simeq \sum_\sigma \sum_{k \in \sigma} S^{*}_{ik} (\nu) S_{jk} (\nu) \, n(T_\sigma,\nu) \, 
    \mathcal{J}_{k}
    \:,
\end{equation}
with
\begin{equation}
    \mathcal{J}_{k}
    \equiv \sum_{m \in \rho (k)} \frac{S_{jm} (\nu)}{S_{jk} (\nu)} \, \gamma_{c, km}^\sigma  \:,
\end{equation}
where $\rho (k)$ is a collection of modes close enough to mode $k$ such that
$\gamma_{c, km}^\sigma$ is non-zero.
The expressions for normalized coherence, both $\gamma_{ij}$ and $\gamma^{(2)}_{ij}$, become identical to the case with completely incoherent sources
if $\mathcal{J}_{k}$ is constant across $k \in \sigma$.\footnote{Strictly speaking, reducing $\mathcal{J}_{k}$ modifies the normalized coherence from the case with complete incoherence. However, these changes in $\mathcal{J}_{k}$ lead to changes in apparent brightness temperature, which can be absorbed into $T_\sigma$, and therefore the formal equivalence for the normalized coherence still holds.}
Whether $\mathcal{J}_{k}$ can be regarded as constant should be evaluated on a case-by-case basis.

For the cases discussed in this paper, we can regard $\mathcal{J}_{k}$ as constant given the following arguments. For aperture radiation,
$k$ and $m$ correspond to modes emitted by infinitesimal sources at $(u_k, v_k)$ and $(u_m, v_m)$, respectively.  Using Eqs.~(\ref{equ:classical_coupling_in_telescope}) and (\ref{equ:thetai_phii_vs_F_xi_yi}), $\mathcal{J}_{k}$ can be written as
\begin{equation}
\begin{split}
    \mathcal{J}_{k}
   & = \iint_{m \in \rho(k)} \mathrm{d}u_m \, \mathrm{d}v_m \, \frac{G_j(u_m, v_m)}{G_j(u_k, v_k)}
   \\
   & \quad \quad \quad \cdot e^{2\pi i [(u_m-u_k)x_j + (v_m-v_k)y_j]/D_{\rm ap} F \lambda} \gamma_{c, km}^\sigma \:,
\end{split}
\end{equation}
with the integrated region $m \in \rho(k)$ being
\[
   s \equiv \sqrt{(u_m-u_k)^2 + (v_m-v_k)^2} \leq R_c \:.
\]
Since $R_c \sim \lambda \ll D_{\rm ap}$, the aperture illumination function varies minimally within the
integrated range and thus $G_j(u_m, v_m) \simeq G_j(u_k, v_k)$.
The source coherence function $\gamma_{c, km}^\sigma$ is isotropic and thus depends only on $s$ as $\gamma_{c, km}^\sigma = \gamma_{c}^\sigma (s)$, leading to
\begin{equation}
  \mathcal{J}_{k} \simeq
  \int_0^{2\pi} \!\! \!\! \dd \phi \int_{0}^{R_c}  \! \! \! \! \! \!
            s \, \dd s
            \, \gamma_{c}^\sigma (s)
  \, e^{2\pi i \cdot s \, d_j \cdot \cos \phi / D_{\rm ap} F \lambda}
   \:,
\end{equation}
with
\[
   d_j \equiv \sqrt{x_j^2 + y_j^2} \:.
\]
Thus, $\mathcal{J}_{k}$ is constant as a function of $k$ and source coherence can therefore be ignored.

For stop radiation, the physical geometry may be more complicated than for the aperture radiation,
and therefore $\mathcal{J}_{k}$ may sometimes vary across the source.  However, because $\mathcal{J}_{k}$ typically
varies slowly compared to $\lambda$, we can divide the source into sections
that are significantly larger than $\lambda$ and in which $\mathcal{J}_{k}$ is effectively constant. Given this setup, we can ignore source coherence by regarding each of these sections as independent sources indexed by $\sigma$ in Eq.~(\ref{equ:mutual_intensity_partial_coherence}).

In summary, for the model optical system presented in Sec.~\ref{sec:optical_model}, the assumption of completely incoherent sources is a good approximation for the context of this paper.

%% file: app_flat_illumination_vs_general.tex
\section{Goodness of Flat-Illumination Approximation}
\label{app:illumination_flat_vs_realistic}

\subsection{Aperture Radiation}
In this section, we show that Eq.~(\ref{equ:flat_illum_approx}) is a good approximation of Eq.~(\ref{equ:mutual_intensity_aperture}) even for a general Gaussian illumination function $G_i(u, v)$.
The key assumption is that the beam can be approximated by Eqs.~(\ref{eq:detector_gaussian_beam}) and (\ref{eq:wf}) and that the pixel spacing satisfies
\begin{equation}
    p_{ij} \geq D_{\rm pix} \:.
\end{equation}

As described in Sec.~\ref{sec:corr_patterns} and Eq.~(\ref{eq:gaussian_illum_func}), we approximate the illumination function as
\begin{equation}
    G_i(u, v) = \frac{1}{\sqrt{\pi \sigma_{\rm ap}^2}}
    \exp \left( - \frac{u^2 + v^2}{2 \sigma_{\rm ap}^2} \right) \: ,
\label{eq:gaussian_illumination_function}    
\end{equation}
and as described in Eq.~(\ref{eq:detector_gaussian_beam}), the circumference of the aperture corresponds to an angle of $\theta \simeq 1/2F$.  It thus follows that
\begin{equation}
    \sigma_{\rm ap} = D_{\rm ap} \frac{w_f}{\sqrt{2} \pi} \frac{F \lambda}{D_{\rm pix}}
    \geq D_{\rm ap} \frac{w_f}{\sqrt{2} \pi} \frac{F \lambda}{p_{ij}} \:,
\end{equation}
assuming a linear mapping from detector-beam angle $(\theta, \phi)$ to aperture position $(u,v)$.

\begin{figure}[tbp]
        \centering
        \includegraphics[width=0.372\textwidth, trim=0.3cm 1.3cm 3cm 3cm, clip]{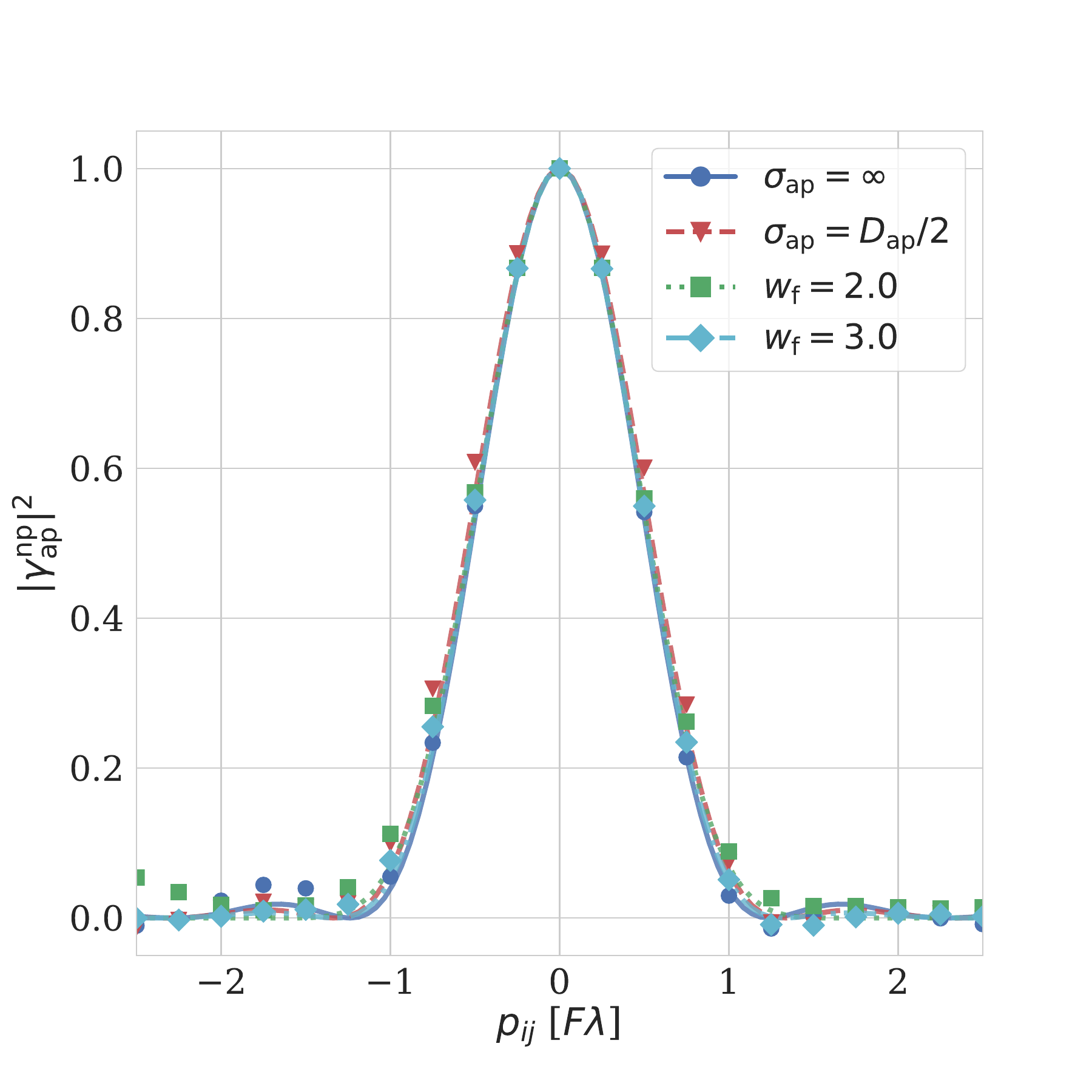}
    \caption{Simulated correlation at the focal plane due to radiation within the aperture for four representative illumination patterns: $\sigma_{\mathrm{ap}} = (\infty, D_{\mathrm{ap}}/2)$, which demonstrates the goodness of the uniform-illumination approximation for Gaussian beams, and $w_f = (2, 3)$, which demonstrates the goodness of the pixel-size-dependent illuminations in Eq.~(\ref{eq:detector_gaussian_beam}). Similarly to Fig.~\ref{fig:hbt_corr_coeff}, the points show the results of Monte Carlo simulations while the lines show the expectation of VCZT.}
    \label{fig:hbt_corr_coeff_variousIllum}
\end{figure}

To evaluate the approximation in Eq.~(\ref{equ:flat_illum_approx}), which corresponds to the limit of $\sigma_{\rm ap} \rightarrow \infty$ or $D_{\rm pix} \rightarrow 0$, we evaluate the opposite limit of $p_{ij} = D_{\rm pix}$ or
\[
     \sigma_{\rm ap,min}(p_{ij}) =  D_{\rm ap} \frac{w_f}{\sqrt{2} \pi} \frac{F \lambda}{p_{ij}} \: ,
\]
where the approximation is its worst. In Fig.~\ref{fig:hbt_corr_coeff}, we show the HBT correlation coefficient $|\gamma(p_{ij})|^2$ calculated using $\sigma_{\rm ap} = \infty$ and $\sigma_{\rm ap} = \sigma_{\rm ap,min}(p_{ij})$ for two typical $w_f$ values, and all curves match well.

\subsection{Stop Radiation}
\begin{figure}[tbp]
        \centering
        \includegraphics[width=0.45\textwidth]{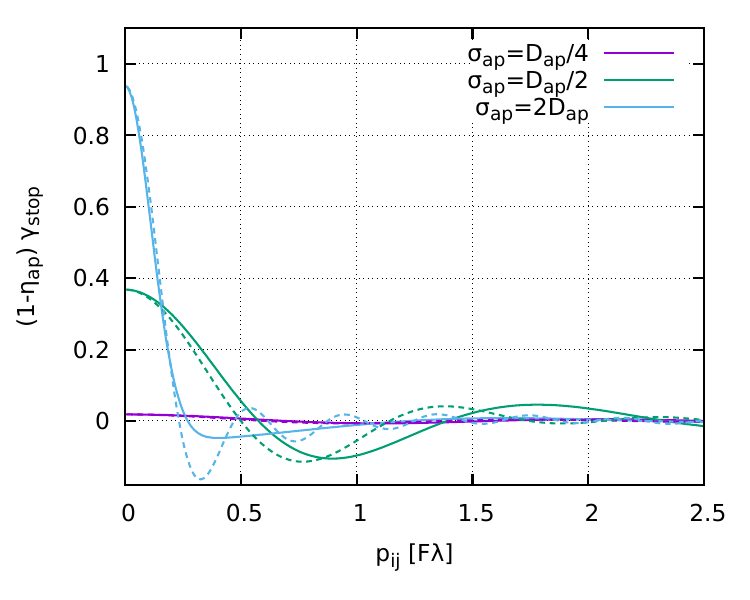}
    \caption{The fractional contribution of the stop radiation, $(1-\eta_{\rm ap}) \, \gamma^{\rm np}_{{\rm stop}, ij}$, to the VCZT coefficient.
    Solid lines show the numerical calculation assuming a Gaussian beam, and dashed lines show the approximation 
    $(1-\eta_{\rm ap}) \, \gamma_{{\rm approx}, ij} (\nu)$.
    }
    \label{fig:stop_radiation_with_various_sigma_ap}
\end{figure}

As Fig.~\ref{fig:pixel_packing_contours} suggests, stop radiation contributes only minorly to the total correlation among pixels. To demonstrate this contribution explicitly, Fig.~\ref{fig:stop_radiation_with_various_sigma_ap} shows $(1-\eta_{\rm ap}) \, \gamma^{\rm np}_{{\rm stop}, ij}$ for various $\sigma_{\rm ap}$ of the Gaussian illumination in Eq.~(\ref{eq:gaussian_illumination_function}).  When $\sigma_{\rm ap} \gg D_{\rm ap}$, the VCZT coherence $\gamma^{\rm np}_{{\rm stop}, ij}$ asymptotes to a Dirac delta function and thus contributes negligibly to the intensity correlation. In the other limit of $\sigma_{\rm ap} \ll D_{\rm ap}$, $(1-\eta_{\rm ap}) = \exp(-D_{\rm ap}^2 / 4 \sigma_{\rm ap}^2)$ asymptotes to zero and the stop becomes irrelevant.

The contribution of stop radiation may become non-negligible, though still small, when $\sigma_{\rm ap}$ is in neither of these limits, or when $\sigma_{\rm ap} \sim D_{\rm ap}/2$.  In this parameter region, an approximation can be obtained by calculating the coherence of a flat-illuminated annulus with width $\sigma_{\rm ap}$
\begin{equation}
  \begin{split}
    & \gamma^{\rm np}_{{\rm stop}, ij} (\nu)
    \\
    & \; = \frac{1}{(1- \eta_{\rm ap}) }
    \iint_{D_{\rm ap}/2 \leq \sqrt{u^2 + v^2}}
    \!\!\!\!\!\!\!\!\!\!\!\!\!\!\!\!\!\!\!\!\!\!\!\!\!\! \dd u \, \dd v \; |G_i(u, v)|^2 \,
    e^{2\pi i \, p_{ij} u /D_{\rm ap} F \lambda}
    \:,
    \\
    & \; \simeq 
    \left\{
       \pi \left(\frac{D_{\rm ap}}{2}+\sigma_{\rm ap}\right)^2
       - \pi \left(\frac{D_{\rm ap}}{2}\right)^2
    \right\}^{-1}
    \\
    & \quad \quad \quad \quad \cdot  \iint_{D_{\rm ap}/2 \leq \sqrt{u^2 + v^2} < D_{\rm ap}/2+ \sigma_{\rm ap}}
    \!\!\!\!\!\!\!\!\!\!\!\!\!\!\!\!\!\!\!\!\!\!\!\!\!\! \dd u \, \dd v \; e^{2\pi i \, p_{ij} u /D_{\rm ap} F \lambda}
    \:, \\
    & \; = \frac{F^2 F'^2}{F^2 - F'^2}
    \left\{ \frac{1}{F'^2}\frac{2 J_1 (\pi p_{ij} / F' \lambda)}{\pi p_{ij} / F' \lambda}
    -  \frac{1}{F^2}\frac{2 J_1 (\pi p_{ij} / F \lambda)}{\pi p_{ij} / F \lambda} \right\}
    \\
    & \; \equiv \gamma_{{\rm approx}, ij} (\nu) \; ,
    \end{split}
  \label{eq:approx_stop_radiation}
\end{equation}
where
\[
   F' \equiv \frac{D_{\rm ap}}{D_{\rm ap} + 2 \sigma_{\rm ap}} F \:.
\]
The dashed lines in Fig.~\ref{fig:stop_radiation_with_various_sigma_ap} show this approximation for a few relevant examples.